\RequirePackage{fix-cm}
\documentclass[twocolumn]{svjour3}
\smartqed  % flush right qed marks, e.g. at end of proof
\usepackage{natbib}
\setcitestyle{authoryear,round}
\usepackage{graphicx}
\usepackage{balance}
\usepackage{multirow}
\usepackage{multicol}
\usepackage{amsmath}
\usepackage{amssymb}
\usepackage{soul}
\usepackage{booktabs}
\usepackage[colorlinks,linkcolor=red,anchorcolor=blue,citecolor=blue,CJKbookmarks=True]{hyperref}

\begin{document}

\title{Probabilistic-based Feature Embedding of 4-D Light Fields for Compressive Imaging and Denoising
}


\author{Xianqiang Lyu         \and
        Junhui Hou %etc.
}


\institute{X. Lyu and J. Hou \at
              Department of Computer Science, City University of Hong Kong, and City University of Hong
Kong Shenzhen Research Institute.\\
             % Fax: +123-45-678910\\
              \email{xianqialv2-c@my.cityu.edu.hk and jh.hou@cityu.edu.hk}           %  \\
           \and
           Manuscript accepted in IJCV. This work was supported in part by the National Key R\&D Program of China No.2022YFE0200300, in part by the Hong Kong Research Grants
Council under Grants 11218121 and 21211518, and 
in part by the Hong Kong Innovation and Technology Fund under Grant MHP/117/21.
        %    second address
}

\date{Received: 5 May 2023 / Accepted: 7 December 2023}
% Accepted in IJCV

\maketitle

\begin{abstract}
\sloppy
The high-dimensional nature of the 4-D light field (LF) poses great challenges in achieving efficient and effective feature embedding, that severely impacts the performance of downstream tasks. To tackle this crucial issue, in contrast to existing methods with empirically-designed architectures, we propose a probabilistic-based feature embedding (PFE), which learns a feature embedding architecture by assembling various low-dimensional convolution patterns in a probability space for fully capturing spatial-angular information. Building upon the proposed PFE, we then leverage the intrinsic linear imaging model of the coded aperture camera to construct a cycle-consistent 4-D LF reconstruction network from coded measurements. Moreover, we incorporate PFE into an iterative optimization framework for 4-D LF denoising. Our extensive experiments demonstrate the significant superiority of our methods on both real-world and synthetic 4-D LF images, both quantitatively and qualitatively, when compared with state-of-the-art methods. The source code will be publicly available at \href{https://github.com/lyuxianqiang/LFCA-CR-NET}{https://github.com/lyuxianqiang/LFCA-CR-NET}.

%\subclass{MSC code1 \and MSC code2 \and more}
\end{abstract}

\begin{figure*}
  \includegraphics[width=1\linewidth]{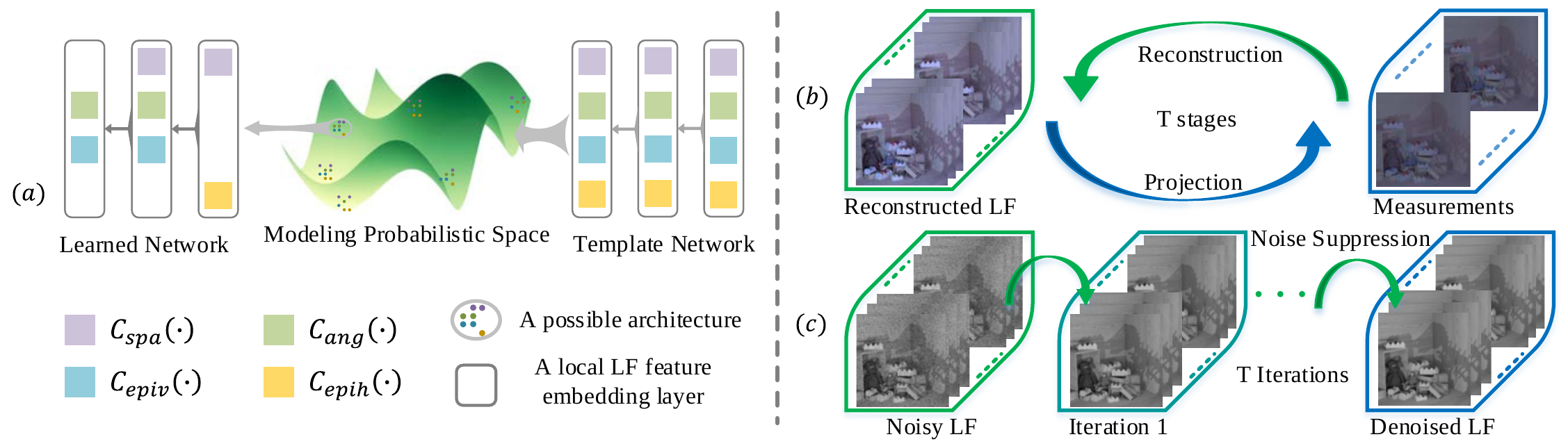}
  \caption{(a) Illustration of probabilistic-based feature embedding. The learned network is derived from the MAP estimation in a probability space using a template network that comprises multiple layers of basic convolution of LF slices, including spatial, angular, vertical EPI, and horizontal EPI slices. Building upon this module, a cycle-consistent framework (b) has been devised for compressive imaging, which involves several iterative stages of projection and reconstruction. During the projection phase, the coded aperture imaging process is simulated. Additionally, we propose an iterative denoising framework (c) comprising multiple iterative noise suppression modules.
  }
  \label{fig:pipeline2}
\end{figure*}

\sloppy
\section{Introduction} \label{intro}
The 4-D light field (LF) records both the spatial and angular information of light rays emanating from the 3-D scene. 
Owing to its ability to capture richer information, the LF has found applications in various fields, such as digital refocusing \cite{ng2005light, ng2006digital}, depth estimation \cite{wanner2013variational,wang2015occlusion,park2017robust,jin2022occlusion}, saliency detection \cite{li2014saliency,wang2019deep,jing2021occlusion}, object recognition \cite{wang20164d}, and segmentation \cite{hog2016light}. 
However, this high-dimensional nature of LF data also presents new challenges, particularly in efficiently and effectively extracting its information and features for diverse applications.

To process the 4-D LF, previous learning-based methods simply apply 2-D or 3-D convolutional filters to the stack of sub-aperture images (SAIs) \cite{inagaki2018learning,vadathya2019unified,shin2018epinet,zhang2019residual}, which has a limited ability to explore the angular relations among SAIs. Some advanced network architectures have been designed to model the LF structure better, such as the 4-D convolutional layer \cite{yeung2018light}, the epipolar plane images (EPIs) convolutional layer \cite{wu2017light,wu2019learning,heber2017neural}, the spatial-angular separable (SAS) convolutional layer \cite{yeung2018fast,Jin_2020_CVPR,guo2020deep,jin2020deep}, the spatial-angular interactive network \cite{wang2020spatial}, the angular deformable convolution \cite{wang2020light}, and the disentangle convolution on macro-pixel image \cite{wang2022disentangling}.
However, as these methods combine the spatial and angular features of the LF data in an intuitive or empirical manner, the optimal strategy to model the LF data still needs to be studied. Inspired by the recent advances of Transformer, several Transformer-based methods \cite{wang2022detail,liang2022light,Liang2023EPIT} have been proposed for many LF processing tasks, which can incorporate the information from all angular views and capture long-range spatial dependencies in each SAI.

In this paper, to embed 4-D LF data efficiently and effectively, we propose a probabilistic-based feature embedding module. As shown in Fig. \ref{fig:pipeline2}, we formulate the problem in a probability space and propose to approximate a maximum posterior distribution (MAP) of a set of carefully defined LF processing events, including both layer-wise spatial-angular feature extraction and network-level feature aggregation. Through droppath from a densely-connected template network, we derive an adaptively probabilistic-based feature embedding, which is sharply contrasted with existing manners that combine spatial and angular features empirically. Building upon the proposed probabilistic-based feature embedding method, we propose two distinct frameworks for the essential tasks of compressive LF imaging and LF image denoising. These tasks are critical in the context of LF data transmission and preprocessing for subsequent applications.

Earlier 4-D LF capturing devices, such as camera array \cite{wilburn2005high} and camera gantry \cite{lfgantry}, are bulky and costly to capture dense LFs. Although more recent commercial LF cameras, such as \cite{Lytro} and \cite{RayTrix}, are more convenient and efficient, the limited sensor resolution results in a trade-off between spatial and angular resolution. Various manners have been proposed to obtain high-quality 4-D LFs (i.e., 4-D LFs with both high spatial and angular resolution), such as spatial super-resolution \cite{wang2020light,Jin_2020_CVPR,wang2022disentangling,van2023light,NTIRE2023-LFSR}, angular super-resolution (or view synthesis) \cite{jin2020deep,chen2022light,yang2023light}, hybrid lenses \cite{wang2016light,jin2023lighthybrid}, and coded aperture imaging \cite{inagaki2018learning,vadathya2019unified,guo2021deep}. Particularly, coded aperture imaging, which encodes the 4-D LF to 2-D coded measurements (CMs) without losing the spatial resolution, demonstrates to be a promising way for high-quality 4-D LF acquisition \cite{wilburn2005high}. However, the bottleneck of this manner lies in the ability of the subsequent reconstruction algorithm, which reconstructs the 4-D LF image from the 2-D CMs.
Although existing deep learning-based 4-D LF reconstruction methods from CMs \cite{inagaki2018learning,vadathya2019unified} have presented significantly better reconstruction quality than conventional ones \cite{babacan2012compressive,marwah2013compressive}, they generally employ plain convolutional networks that are purely data-driven without taking the observation model into account.
More recently, \cite{guo2020deep} proposed an unrolling-based method to link the observation model of coded apertures and deep learning elegantly and improve the LF reconstruction quality significantly.
In this paper, we propose a physically interpretable framework incorporated with the probabilistic-based feature embedding to reconstruct the 4-D LF from 2-D CMs. Specifically, based on the intrinsic linear property of the observation model, we propose a cycle-consistent reconstruction network (CR-Net), which reconstructs the LF in a progressive manner by gradually eliminating the residuals between the back-projected CMs from the reconstructed LF and input CMs. 

Furthermore, we exploit the probabilistic-based feature embedding approach in the context of LF denoising, which is a fundamental and pressing task for various subsequent LF applications, including but not limited to depth estimation, object recognition, and low light enhancement \cite{lamba2020harnessing}. Similar to conventional 2-D images, noise such as thermal and shot noise can corrupt the captured LF data during the acquisition process. However, the higher dimensionality and unique geometric structures of LF data present significant challenges for extending existing low-dimensional signal denoising methods, e.g., 2-D single image denoiser \cite{zhang2017beyond} or 3-D video denoiser \cite{tassano2020fastdvdnet}, to the 4-D LF domain.
The state-of-the-art deep learning-based method \cite{guo2021deep} considered the LF denoising as a special case of coded aperture LF imaging with the degradation matrix being an identity matrix. The extended pipeline to LF denoising does not fully exploit the characteristics of LF denoising. By sampling analysis for LF noise suppression, we propose an iterative optimization framework for LF denoising with a carefully designed noise suppression module, in which the high-dimensional characteristic of LF denoising can be thoroughly explored. Experimental results demonstrate that the proposed methods outperform state-of-the-art methods to a significant extent. 

In summary, the main contributions of this paper are three-fold:
\begin{itemize}
    \item a probabilistic-based feature embedding method for 4-D LFs;
    \item a novel framework for coded aperture-based 4-D LF reconstruction; and
    \item a novel denoising framework for 4-D LF images. 
\end{itemize}

The rest of this paper is organized as follows. Section \ref{sec2} reviews related works. Section \ref{sec3} presents the proposed probabilistic-based feature embedding. In Section \ref{sec4}, we present the cycle-consistent network for compressive 4-D LF imaging. In Section \ref{sec5}, we present the iterative optimization network for 4-D LF denoising. In Section \ref{sec6}, extensive experiments are carried out to evaluate our framework on LF coded aperture imaging, and LF denoising. Finally, Section \ref{sec7} concludes this paper.

\begin{table*}
\centering
\label{tab:summary}
\caption{Summary of feature extraction strategies used for learning-based LF spatial super-resolution, LF angular super-resolution (or view synthesis), coded aperture-based compressive imaging for LF reconstruction, and LF denoising.}
\begin{tabular}{ccccc}
 \toprule[1.2pt]
Task & Method & Feature Embedding & Processing 4-D information      \\ \hline
\multicolumn{1}{c}{\multirow{9}{*}{Spatial Super-resolution}} 
& \cite{yoon2015learning} & 2-D Conv + SAI stacks   & No                 \\
& \cite{chen2022light}    & 4-D EPI + PSAIB     &   Yes       \\

& \cite{wang2020spatial}    & SAS Interaction    &   Yes   \\
& \cite{wang2020light}    & Angular Deformable Conv    &   Yes   \\
& \cite{wang2022disentangling}    & 2-D Disentangling Conv    &   Yes   \\
& \cite{van2023light}     & 2-D Conv + HFEM    &    Yes \\ 

& \cite{wang2022detail}     & SAS + Transformer    &    Yes \\
& \cite{liang2022light}     & SAS + Transformer    &    Yes \\ 
& \cite{Liang2023EPIT}     & SA Correlation + Transformer    &    Yes \\ \hline

\multicolumn{1}{c}{\multirow{9}{*}{Angular Super-resolution}} 
& \cite{wu2017light}      & 2-D Conv + EPI stacks   & No             \\ 
& \cite{wang2018end}      & 3-D Conv + EPI stacks   & No            \\ 
& \cite{yeung2018fast}    & SAS Conv   &   Yes       \\ 
& \cite{jin2020learning}  & SAS Conv    &   Yes           \\
& \cite{jin2020deep}      & SAS Conv    &   Yes       \\ 
& \cite{chen2022light}    & 4-D EPI + PSAIB     &   Yes       \\
& \cite{wang2022disentangling}    & 2-D Disentangling Conv    &   Yes   \\
& \cite{van2023light}     & 2-D Conv + HFEM    &    Yes \\
& \cite{yang2023light}    & 4-D Conv + 4-D Deconv    &    Yes \\  \hline

\multicolumn{1}{c}{\multirow{4}{*}{Compressive Reconstruction}} 
& \cite{gupta2017compressive} & 4-D Conv + MLP  &    Yes              \\
& \cite{inagaki2018learning}  & 2-D Conv + CMs  &    No        \\ 
& \cite{vadathya2019unified}  & 2-D Conv + CMs  &    No         \\ 
& \cite{guo2021deep}          & SAS Conv        &  Yes       \\ \hline
\multicolumn{1}{c}{\multirow{3}{*}{Denoising}} 
& \cite{chen2018light}   & 2-D Conv + SAI stacks  &   No                \\
& \cite{guo2021deep}     & SAS Conv       &   Yes            \\ 
& \cite{wang2023multi}     & 2-D Conv + SAI stacks  &   No            \\ \bottomrule[1.2pt]
\end{tabular}
\end{table*}

\section{Related Work} \label{sec2}
\subsection{Feature Embedding of 4-D LFs} 
Effective and efficient feature embedding constitutes a fundamental module of diverse learning-based LF tasks and bears a direct relationship to the ultimate performance of the model. Table \ref{tab:summary} enumerates the feature extraction techniques utilized in various associated LF tasks, thereby facilitating an intuitive comprehension of the evolution of feature extraction methodologies. Some methods process the 4-D LF from its low-dimensional representation, i.e., 2-D SAI \cite{yoon2015learning,chen2018light,wang2023multi}, 2-D CMs \cite{inagaki2018learning,vadathya2019unified}, 2-D EPI \cite{wu2017light}, and 3-D EPI \cite{wang2018end} volume, which cannot comprehensively explore the distribution of the high-dimensional data. To comprehensively explore the characteristics of high-dimensional LF, \cite{gupta2017compressive} employed 4-D convolution and MLP to process the 4-D data. Additionally, \cite{yang2023light} devised LF-ResBlock, which is made up of 4-D convolution and deconvolution, to facilitate angular super-resolution. 
In addition to this simple idea, \cite{yeung2018fast} proposed the SAS convolutional layer that can efficiently process the 4-D LF by sequentially conducting 2-D spatial and angular convolutions. This SAS convolutional layer was employed in several LF tasks, such as angular super-resolution \cite{jin2020deep}, compressive reconstruction \cite{guo2020deep}, and LF denoising \cite{guo2021deep}. \cite{chen2022light} extended the SAS convolutional to parallel spatial-angular integration blocks (PSAIBs), achieving dense interaction between the spatial and angular domain in a two-stream fashion. \cite{wang2022disentangling} proposed a disentangling mechanism in the 2-D macro-pixel image to extract spatial, angular, and EPI features. However, the empirical combination of spatial, angular, and EPI features may limit the quality of LF reconstruction. Thus, several methods are proposed to combine the high-dimensional feature in some smarter ways \cite{lyu2021learning,van2023light}. For example, \cite{van2023light} employed a hybrid feature extraction module (HFEM) that operates on all three 2-D subspaces (i.e., spatial, angular, and EPI) of the LF image. Besides these CNN-based methods, many Transformer-based methods \cite{wang2022detail,liang2022light,Liang2023EPIT} have been proposed for different LF processing tasks.

\subsection{Coded Aperture-based 4-D LF Reconstruction} 
Many conventional methods \cite{liang2008programmable,ashok2010compressive,babacan2012compressive,marwah2013compressive} have been proposed to reconstruct the LF from CMs.
\cite{liang2008programmable} proposed a method to reconstruct the LF from CMs captured by the programmable aperture system. However, the method requires that the number of measurements is equal to the angular resolution of the reconstructed LF. \cite{ashok2010compressive} and \cite{babacan2012compressive} made efforts to reduce the number of required measurements by exploiting the spatial-angular correlations, and employing the hierarchical Bayesian model, respectively.  \cite{marwah2013compressive} proposed an LF reconstruction method through the perspective of compressive sensing that can recover the LF from a single CM by the trained overcomplete dictionary. Limited by the representation ability of the conventional mathematical models and the overcomplete dictionary, the quality of reconstructed LFs of these methods is very limited. With the popularization of deep learning, several deep learning-based methods have been proposed for CM-based LF reconstruction. \cite{gupta2017compressive} presented a two-branch network for compressive LF reconstruction. \cite{inagaki2018learning} used the auto-encoder architecture to encode and decode the LF in an end-to-end manner. However, the LF reconstruction module in these methods is designed from a data-driven perspective and suffers from limited LF reconstruction quality.  \cite{vadathya2019unified} proposed to estimate disparity maps from CMs and the predicted central SAI, and then warp the central SAI to other ones by using these disparity maps. The performance of this method is affected by the accuracy of the disparity estimation. 
\cite{guo2020deep} proposed a deep unrolling-based method to reconstruct the LF from CMs in a more interpretable way, which significantly improves the LF reconstruction quality. 

\begin{figure}[h]
  \centering
  \includegraphics[width=0.95\linewidth]{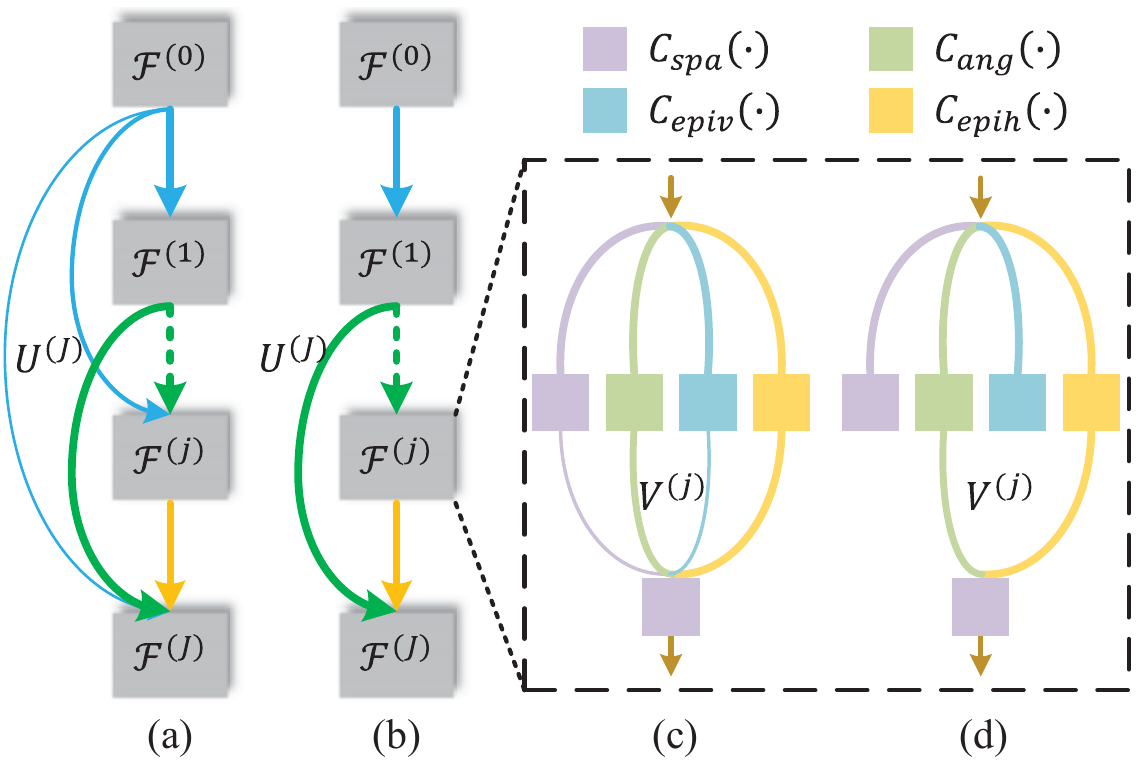}
  \caption{Illustration of network-level feature aggregation and feature embedding unit used in our network. (a,c) The network with all potential paths, namely the template network.
  (b,d) One possible architecture from the posterior distribution $\mathcal{Q}(\mathcal{W}| \mathbb{D})$.
  }
  \label{fig:IJCV_pfe}
\end{figure}

\subsection{4-D LF Image Denoising} 
Since an LF can be represented as a 2-D array of 2-D images, some conventional methods extend image or video denoising algorithms to LFs. For example, the LFBM5D filter \cite{alain2017light} extended the image denoiser BM3D \cite{dabov2007image} to 5-D patches by considering the redundancies in the 2-D angular patch of the LF. In \cite{li2013joint}, a two-stage framework was proposed to separately suppress the noise of horizontal and vertical EPIs by employing an image denoiser proposed in \cite{beck2009fast}. Furthermore, \cite{sepas2016light} utilized the video denoiser \cite{maggioni2012video} to process EPI sequences. From a frequency domain perspective, conducting filtering is an efficient way to segregate the LF signal from noise. \cite{dansereau2013light} explored the characteristics of LFs in the spatial frequency domain and utilized the 4-D hyperfan filter to accomplish LF denoising. In \cite{premaratne2020real}, a real-time LF denoising method was proposed using a novel 4-D hyperfan filter, which is approximated in the 4-D mixed-domain using a 2-D circular filter and 2-D parallelogram filters.
More recently, \cite{tomita2022denoising} proposed a denoising method for multi-view images based on the short-time discrete Fourier transform (ST-DFT). The proposed denoising method first transforms noisy multi-view images into the ST-DFT domain, and then noisy ST-DFT coefficients are denoised by soft thresholding derived from the proposed multi-block Laplacian model.

Recently, some deep learning-based LF denoising methods were proposed, which denoise the LF by training with a large amount of LF data. \cite{chen2018light} proposed a deep learning-based framework for LF denoising with two sequential CNNs. The first network creates the structural parallax details, and the second restores the view-dependent local energies. 
In \cite{guo2021deep}, their deep spatial-angular regularization framework was extended to LF denoising, which implicitly and comprehensively explores the signal distribution based on the observation model of noisy LF images. 
\cite{wang2023multi} utilized the multiple stream progressive restoration network framework and averaged view stacks to LF denoising.
These learning-based methods have demonstrated excellent performance for light field denoising.

\section{Proposed Probabilistic-based Feature Embedding} \label{sec3}
To comprehensively explore the spatial-angular information of 4-D LFs, high-dimensional convolution is an intuitive choice that has demonstrated its effectiveness \cite{wang2018end}. However, compared to 2-D convolution, it significantly increases the number of parameters, which may cause overfitting and consume significant computational resources. By analogy with the approximation of a high-dimensional filter with multiple low-dimensional filters in the field of signal processing, some researchers have proposed to apply convolutions separately on the spatial \cite{yoon2015learning}, angular \cite{yeung2018light}, or EPI \cite{wu2017light} domains. However, it is laborious and inconvenient to design an optimal LF feature embedding module manually. Furthermore, the aggregations of different layers also play a crucial role during LF reconstruction. Inspired by the current success of neural architectural search \cite{liu2019darts,wang2020you}, we introduce an \textit{adaptive probabilistic-based feature embedding module} that builds a probabilistic model to locally choose the LF feature embedding patterns and globally optimize the aggregation patterns of different layers. Thus, both the network architecture and corresponding weights are optimized to achieve efficient and effective LF reconstruction.

To better demonstrate the possible events during LF feature embedding, we first construct the modeling probabilistic space over each probabilistic-based feature embedding module $\mathcal{D}^{(t)}(\cdot) (0\leq t \leq T-1)$. Denote the global probability space of LF-processing CNNs with
$\mathcal{W} = \{ (\mathbf{U}^{(1)},\mathcal{F}^{(1)}) \cdots (\mathbf{U}^{(j)},\mathcal{F}^{(j)}) \cdots  (\mathbf{U}^{(J)},\mathcal{F}^{(J)})\}$, where a binary vector 
$\mathbf{U}^{(j)} \in \{\mathbf{0},\mathbf{1}\}^{j-1}$ with length $j-1$ indicates whether the features from previous $j-1$ layers are used, and $\mathcal{F}^{(j)}$ denotes a feature embedding unit to learn high-level LF embeddings. According to the characteristics of LF images, we further introduce four 2-D convolutional patterns for local LF feature embedding, \textit{i.e.}, $\mathcal{F}^{(j)} = \{ \mathbf{V}^{(j)}, \mathcal{C}_{spa}^{(j)}(\cdot), \mathcal{C}_{ang}^{(j)}(\cdot), \mathcal{C}_{epih}^{(j)}(\cdot), \mathcal{C}_{epiv}^{(j)}(\cdot) \}$, where $\mathcal{C}_{spa}^{(j)}(\cdot)$, $\mathcal{C}_{ang}^{(j)}(\cdot)$, $\mathcal{C}_{epiv}^{(j)}(\cdot)$, and $\mathcal{C}_{epih}^{(j)}(\cdot)$ denote the 2-D convolutional layers on spatial, angular, vertical EPI, and horizontal EPI domains, respectively; $\mathbf{V}^{(j)} \in \{\mathbf{0},\mathbf{1}\}^{4}$ is a binary vector with length of 4 for selecting different convolution patterns. Then we could drive a deep neural network by solving the maximum a posterior (MAP) estimation problem:
\begin{align}
    \widehat{\mathcal{W}} = \mathop{\arg\max}_{\mathcal{W}} \mathcal{Q}\left(\mathcal{W} | \mathbb{D}\right),
    \label{con:eqution4}
\end{align}
where $\mathbb{D} = \{\mathbf{I}, \mathbf{L}\}$ indicates the data distributions. 

\begin{figure*}
    \centering
    \includegraphics[width=0.95\textwidth]{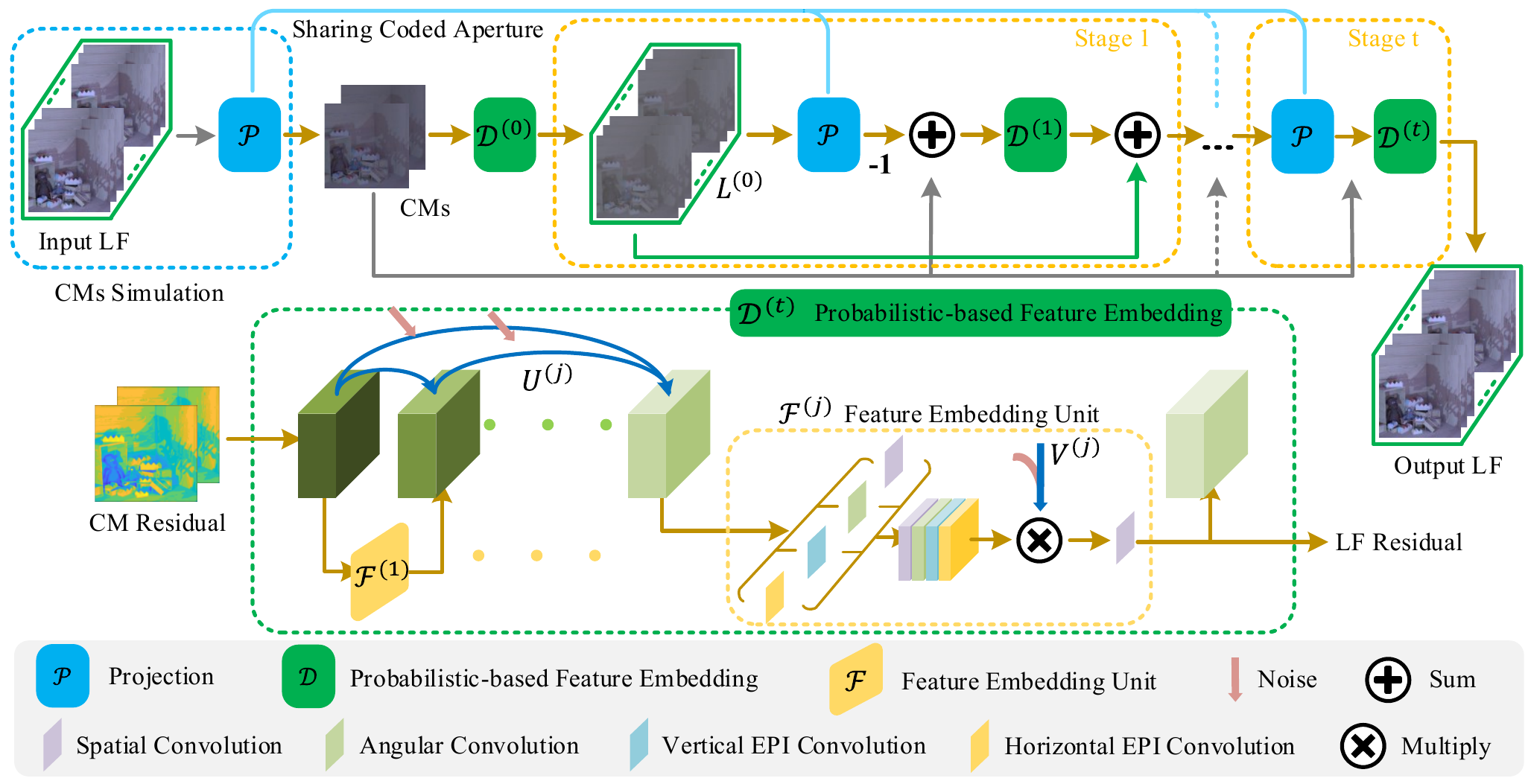}
    \caption{Illustration of the proposed framework named CR-Net for reconstructing 4-D LF images from 2-D measurements by a coded-aperture camera. CR-Net consists of a "CMs Simulation" procedure and the proposed cycle-consistent reconstruction network.
    }
    \label{fig:my_network}
\end{figure*}

To solve Eq. (\ref{con:eqution4}), we first approximate the posterior distribution $\mathcal{Q}\left(\mathcal{W} | \mathbb{D}\right)$. 
According to \cite{gal2016dropout}, embedding dropout into CNNs could objectively minimize the Kullback–Leibler divergence between an approximation distribution and posterior deep Gaussian process \cite{damianou2013deep}. Thus, we could approximate the posterior distribution via training a template network with binary masks that follow learnable independent Bernoulli distributions (we name the network with all $\mathbf{U}^{(j)}$ and $\mathbf{V}^{(j)}$ set to $\mathbf{1}$ as the template network). As shown in Fig. \ref{fig:IJCV_pfe}, both the path for feature aggregation ($\mathbf{U}^{(j)}$) and local feature embedding pattern ($\mathbf{V}^{(j)}$) are replaced by masks of logits $\epsilon \sim \mathcal{B}(p)$, where $\mathcal{B}(p)$ denotes the Bernoulli distribution with probability $p$. However, the classical sampling process makes it hard to manage a differentiable linkage between the sampling results and the probability. Besides, such dense aggregations in CNNs result in a huge number of feature embeddings. Thus, we also need to explore an efficient and effective way of aggregating those features masked by binary logits. In what follows, we detailedly discuss those two aspects.

To obtain a differentiable sampling manner of logits $\epsilon$, we use the Gumbel-softmax \cite{jang2017categorical} to relax the discrete Bernoulli distribution to continuous space. Mathematically,
we formulate this process as
\begin{align}
    \small
    \mathcal{M}(p) = \textsf{Sigmoid} \{ & \frac{1}{\tau }(\log p - \log (1 - p) + \log (\log ({r_1})) \nonumber \\
    & - \log (\log ({r_2}))) \},
    \label{con:eqution5}
\end{align}
where $r_1$ and $r_2$ are random noises with standard uniform distribution in the range of $[0,1]$; $p$ is a learnable parameter encoding the probability of aggregations in the neural network; $\tau > 0$ is a temperature that controls the similarity between $\mathcal{M}(p)$ and $\mathcal{B}(1-p)$, \textit{i.e.},
as $\tau \to 0$, the distribution of $\mathcal{M}(p)$ approaches $\mathcal{B}(1-p)$; while as $\tau \to \infty$, $\mathcal{M}(p)$ becomes a uniform distribution.

To aggregate the features efficiently and effectively, we design the network architecture at both the network and layer levels.
According to  Eq. (\ref{con:eqution5}), we approximate the discrete variable $\mathbf{U}^{(j)}$ by applying Gumbel-softmax $\mathcal{M}(\cdot)$ to continuous learnable variables $\widetilde{\mathbf{U}}^{(j)}$. Thus, we could formulate network-level feature aggregation as
\begin{align}
&\widetilde {\mathbf{H}}^{(j)} = { \mathcal{C}_{1 \times 1}}\left( {  { \mathbf{T}^{(0,j)} , \cdots, \mathbf{T}^{(j-1,j)}}} \right),\nonumber\\
&{\rm with} ~~ \mathbf{T}^{(k,j)} = {\mathbf{H}}^{(k)} \times \mathcal{M}\left(\widetilde{\mathbf{U}}^{(j)}(k)\right),
    \label{con:eqution6}
\end{align}
where $\widetilde{\mathbf{H}}^{(j)} ~~ (1 \leq j \leq J)$ denotes the aggregated feature which would be fed into $\mathcal{F}^{(j)}$; $\mathbf{H}^{(k)} ~~ (1 \leq k \leq j-1)$ represents the feature from the ${k}$-th embedding unit $\mathcal{F}^{(k)}$; $\mathbf{H}^{(0)}$ denotes an LF embedding extracted from the  input of $\mathcal{D}^{(t)}(\cdot)$ by a single linear convolutional layer;
$\widetilde{\mathbf{U}}^{(j)}(k)$ indicates the ${k}$-th element of the vector $\widetilde{\mathbf{U}}^{(j)}$, which is random initialized and kept in a range of $[0,1]$ according to its meaning of the sampling probability of $\mathbf{U}^{(j)}$;
and $\mathcal{C}_{1 \times 1}(\cdot)$ represents a $1 \times 1$ kernel to compress the feature embedding and activate them with $ReLU$.

In analogy to the network level, we also introduce the continuous learnable weights $\widetilde{\mathbf{V}}^{(j)}$ with Gumbel-softmax to approximate the Bernoulli distribution of $\mathbf{V}^{(j)}$ in each feature embedding unit, as
\begin{align}
&{\mathbf{H}^{(j + 1)}} = {\mathcal{C}_{spa}}\left( {{{\mathcal{C}}_{1 \times 1}}\left(
\mathbf{O}^{(1,j)},\cdots,\mathbf{O}^{(4,j)}   \right)} \right),\nonumber \\
&{\rm with} ~~ \mathbf{O}^{(l,j)} = \mathcal{C}_{l}^{(j)}\left(\widetilde {\mathbf{H}}^{(j)}\right) \times \mathcal{M}\left(\widetilde{\mathbf{V}}^{(j)}(l)\right),
    \label{con:eqution7}
\end{align}
where $\mathcal{C}_{l}^{(j)}(\cdot) ~~(l \in \{ 1,2,3,4 \})$ indicates the previously designed one of the four convolution patterns in $\mathcal{F}^{(j)}$, which are $\mathcal{C}_{spa}^{(j)}(\cdot)$, $\mathcal{C}_{ang}^{(j)}(\cdot)$, $\mathcal{C}_{epih}^{(j)}(\cdot)$, and $\mathcal{C}_{epiv}^{(j)}(\cdot)$, respectively. After applying a $1\times1$ convolutional layer to compress the embedding, we adopt a further spatial convolution after fusing the feature from different patterns, due to that spatial convolution plays an essential role during the feature embedding process. 

Training such a masked template neural network results in a posterior distribution $\mathcal{Q}(\mathcal{W}|\mathbb{D})$.
Meanwhile, through droppath with lower probability, we could finally derive a neural network with maximum posterior probability (see Fig. \ref{fig:plot-network} for the detailed network architecture).

\section{Proposed Coded Aperture-based 4-D LF Reconstruction}
\label{sec4}
\textbf{Problem Statement}. Denote by $\mathbf{L}(u,v,x,y) \in \mathbb{R}^{M \times N \times H \times W}$ the 4-D LF, where $\{(u,v) | u\in [1,M],v\in [1,N]\}$ and $\{(x,y) | x\in [1,H],y\in [1,W]\}$ are angular and spatial positions, respectively. Then, the $i$-th 2-D CM, denoted as $\mathbf{I}_i\in \mathbb{R}^{H\times W}$, captured by the coded aperture camera can be formulated as:
\begin{equation}
\label{con:eqution1}
\mathbf{I}_i(x,y)=\sum_{u=1}^{M}\sum_{v=1}^{N}a_i(u,v)\mathbf{L}(u,v,x,y),\\
\end{equation}
where $a_i(u,v)\in [0,1]$ is the transmittance at aperture position $(u,v)$ for the $i$-th acquisition. Following  \cite{inagaki2018learning}, we simulate the imaging process in Eq. (\ref{con:eqution1}) to make the coded aperture jointly learned with the subsequent reconstruction. Specifically, shown as the "CMs Simulation" procedure in Fig. \ref{fig:my_network}, we utilize a convolutional layer, denoted as $\mathcal{P}(\cdot)$, with the input of the 4-D LF to generate $S$ CMs, denoted as $\mathbf{I}\in \mathbb{R}^{S \times H\times W}$.
To retrieve $\mathbf{L}$ from $\mathbf{I}$ utilizing a deep learning-based methodology that is predicated on probabilistic-based feature embedding, it is imperative to devise an efficient framework that is tailored to the characteristics of LF compressive imaging. This constitutes a crucial concern for compressive LF reconstruction.\\

\begin{figure*}
    \centering
    \includegraphics[width=0.95\textwidth]{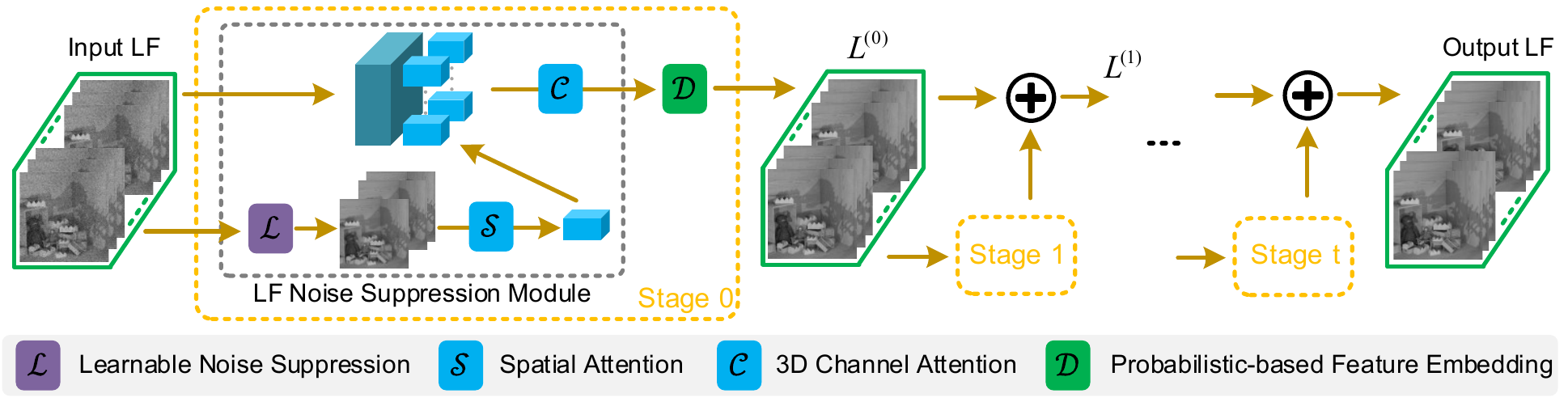}
    \caption{Illustration of the process of the proposed 4-D LF denoising method, namely DN-Net.
    }
    \label{fig:dn_network}
\end{figure*}

\noindent \textbf{Our Solution}.
 The coded aperture model described in Eq. (\ref{con:eqution1}) indicates the cycle consistency between the LF and CMs, i.e., a well-reconstructed LF image could be accurately projected to input CMs. 
Based on this observation,  we propose a
\textit{ cycle-consistent reconstruction framework}, which progressively refines the reconstructed LF via iteratively projecting the reconstructed LF images to pseudo-CMs, then learning a correction map from the differences between measured CMs and pseudo-CMs.
Specifically, as illustrated in Fig. \ref{fig:my_network}, the proposed CR-Net basically consists of two modules, i.e., coarse estimation and cycle-consistent LF refinement.\\

\noindent\textit{Coarse Estimation.} We first learn a coarse estimation $\mathbf{L}^{(0)}$ of the LF from CMs, expressed as:
\begin{equation}
{\mathbf{L}^{(0)}} = \mathcal{D}^{(0)}\left(\mathbf{I}\right),
    \label{con:eqution2}
\end{equation}
where $\mathcal{D}^{(0)}(\cdot)$ denotes a probabilistic-based feature embedding module for the coarse LF estimation.\\

\noindent \textit{Cycle-consistent LF Refinement.} We iteratively learn the LF refinement from the residuals between pseudo-CMs and measured CMs, as:  
\begin{equation}
{\mathbf{L}^{(t+1)}} = \mathcal{D}^{(t)}\left(\mathbf{I}- \mathcal{P}\left(\mathbf{L}^{(t)}\right)\right) + \mathbf{L}^{(t)},
    \label{con:eqution3}
\end{equation}
where $\mathcal{P}(\cdot)$ represents the projection process, which is the same as the convolutional layer used for aperture learning, and  
$\mathcal{D}^{(t)}(\cdot) (t=1,\cdots,T-1)$ 
indicates the probabilistic-based feature embedding module for LF refinement. After iteratively refining the LF images, we could get the final LF estimation, denoted as $\mathbf{L}^{(T)}$.

\begin{figure}[t]
  \centering
  \includegraphics[width=0.95\linewidth]{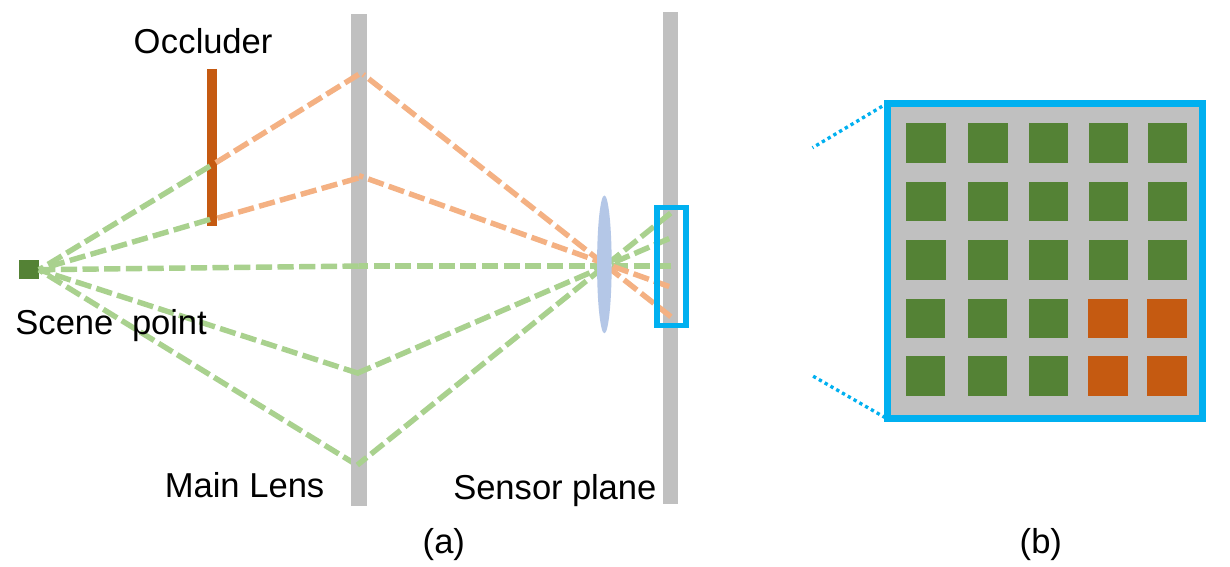}
  \caption{Illustration of imaging model when the occlusion occurs. (a) The imaging process influenced by the occluder. (b) Corresponding angular patch by fixed spatial positions.}
  \label{fig:dn-f1}
\end{figure}

\section{Proposed 4-D LF Image Denoising} 
\label{sec5}
\textbf{Problem Statement}. 
Let $\mathbf{L}_{noisy} \in \mathbb{R}^{M \times N \times H \times W}$ denote an observed noisy 4-D LF, with an observation model $\mathbf{L}_{noisy} = \mathbf{L} + \mathbf{N}$, where $\mathbf{L} \in \mathbb{R}^{M \times N \times H \times W}$ and $\mathbf{N}\sim \mathcal{N}(\mu ,{\sigma ^2})$ are the noise-free LF and additive white Gaussian noise, respectively. Our objective is to restore $\mathbf{L}$ from 
$\mathbf{L}_{noisy}$.

Assuming a Lambertian model, the light ray emitted by a target scene is recorded by different SAIs of a 4-D LF from various perspectives, and the light ray's intensity should be consistent across the SAIs, denoted by $\mathbf{L}_{noisy}^i \in \mathbb{R}^{ H \times W}$. In the presence of additive white Gaussian noise, the observed light intensity from each SAI conforms to a Gaussian distribution, i.e., $l_{noisy}(u,v,x,y) \sim \mathcal{N}(l(u,v,x,y),{\sigma ^2})$, where $l_{noisy}$ and $l$ correspond to the intensity of the noisy and noise-free light, respectively. For simplicity, we analyze a group of light rays corresponding to a single scene/object point.
As shown in Fig. \ref{fig:dn-f1}, we obtain a sampling set of light rays in different SAIs, denoted as $l_{noisy}^{all} = \left\{ {l_{noisy}^1,...,l_{noisy}^Q} \right\}$, where $Q \leq M\times N $ independent and identically distributed samples.
We estimate the noise-free light intensity $l$ using maximum likelihood estimation (MLE) as follows:
\begin{equation}
    {l_{MLE}} = \frac{1}{Q}\sum\limits_{i = 1}^Q {l_{noisy}^i}
    \label{dn:eqution1}
\end{equation}

\noindent \textbf{Our Solution}. 
In the presence of occlusions, the number of samples in the sampling set may be less than the angular number, i.e., $Q < M \times N$. Directly applying Eq. \eqref{dn:eqution1} to all SAIs can lead to the loss of high-frequency details in the noise suppression result due to occlusion and parallax information. As shown in Fig. \ref{fig:dn_network}, we propose an iteratively optimized network for LF denoising. The proposed network leverages probabilistic-based feature embedding and introduces an LF noise suppression module that consists of three particular parts to address the aforementioned blur effect, i.e., learning-based noise suppression, spatial attention for the noise suppression feature, and 3-D channel attention for the combined feature. The subsequent iterative stages predict the residual for denoised LF refinement.\\

\noindent \textit{Learning-based Noise Suppression.} Extending Eq. \eqref{dn:eqution1} to angular averaging of all SAIs can suppress noise and provide guidance for each SAI denoising. However, the occlusion relation can cause high-frequency details to be lost by mixing scene and occluder information. To address this issue, we propose a learning-based noise suppression denoted as $\mathcal{L}(\mathbf{L}_{noisy})$, which concatenates the angular averaging result with seven adaptive angular fusion results.\\

\noindent \textit{Spatial Attention for Noise Suppression Feature.}
The parallax structure of the LF can also result in blurry noise suppression outcomes. Therefore, we incorporated a spatial attention module to process the features of the noise suppression $\mathcal{L}(\mathbf{L}{noisy})$ and reduce the impact of the blurred region. The spatial attention module is defined as follows:
\begin{equation}
\mathcal{F}^S = \texttt{SA}(\mathcal{F}(\mathcal{L}(\mathbf{L}_{noisy})))
\label{dn:eqution2}
\end{equation}
Here, $\mathcal{F}$ and $\mathcal{F}^S$ represent the feature of $\mathcal{L}(\mathbf{L}_{noisy})$ and the output of the spatial attention module, respectively. The function $\texttt{SA}(\cdot)$ denotes the spatial attention processing.\\

\noindent \textit{3-D Channel Attention for the Combined Feature.} We utilize the feature $\mathcal{F}^S$ to guide the LF denoising by concatenating it with each SAI's feature $\mathcal{F}(\mathbf{L}_{noisy}^i)$ and then applying 3-D channel attention to filter out the effective channels. This is expressed as:
\begin{equation}
    \mathcal{F}^C = \texttt{CA}({Cat}_{i=1}^{M \times N}(\mathcal{F}(\mathbf{L}_{noisy}^i),\mathcal{F}^S)
    \label{dn:eqution3}
\end{equation}
Here, $\texttt{CA}(\cdot)$ represents the 3-D channel attention processing, and $\mathcal{F}^C$ denotes the output of the 3-D channel attention module, which is then processed by the subsequent probabilistic-based feature embedding module.

\begin{figure*}
    \centering
    \includegraphics[width=0.95\textwidth]{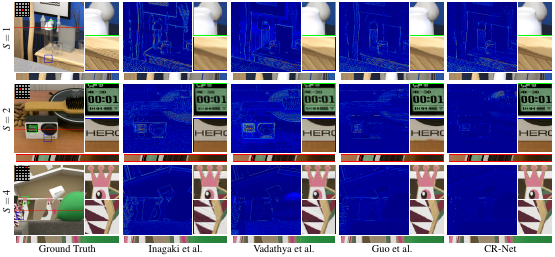}
    \caption{Visual comparisons of reconstructed LFs from different methods under tasks $S=1$, $S=2$, and $S=4$, over the $\textit{HCI+Inria}$ dataset. The error maps of SAIs and zoomed-selected regions are shown for better visualization.}
    \label{fig:IJCV_CA_synf1}
\end{figure*}

\begin{table*}
\centering
\caption{Quantitative comparisons (PSNR/SSIM) of different methods over the test LFs in \textit{Lytro} and \textit{HCI+Inria} datasets under three tasks. The number of network parameters (M) of different methods is also provided.}
\label{tab:table1}
  \begin{tabular}{c|c|ccc|c}
    \toprule[1.2pt]
    Dataset & Method &S = 1  &S = 2 &S = 4 & \# Parameters \\\hline
    \multirow{4}{*}{\textit{Lytro}} & \cite{inagaki2018learning} & 32.11/0.910 & 39.33/0.971 & 40.17/0.975 & 0.69\\
    & \cite{vadathya2019unified} & \textbf{35.62}/\textbf{0.964} & 38.46/0.979 & 39.27/0.985 & 5.14 \\
    & \cite{guo2021deep} & 34.39/0.941 & 41.66/0.984  & 43.19/0.988 & 3.33\\
    & CR-Net & 34.43/0.938 & \textbf{42.54}/\textbf{0.985} & \textbf{45.18}/\textbf{0.990} & 1.75\\\hline
    \multirow{4}{*}{\textit{HCI+Inria}} &  \cite{inagaki2018learning} & 30.60/0.857 &  34.91/0.934 & 35.4/0.928 & 0.69\\
    & \cite{vadathya2019unified} & 30.82/0.866 & 32.69/0.925 & 36.46/0.950 & 5.14 \\
    & \cite{guo2021deep} & 30.72/0.873 & 38.55/0.963 & 42.41/0.976 & 3.33\\
    & CR-Net & \textbf{32.17}/\textbf{0.885} & \textbf{42.49}/\textbf{0.973} & \textbf{46.88}/\textbf{0.983} & 1.75\\
  \bottomrule[1.2pt]
\end{tabular}
\end{table*}

\section{Experiments} \label{sec6}
\subsection{Experiment Settings}
\textbf{Datasets.}
We conducted experiments on both simulated and real CMs in compressive LF imaging. Specifically, we used the same datasets as those used in \cite{guo2020deep} for simulated CMs, which include two LF datasets named \textit{Lytro} and \textit{HCI+Inria}. The \textit{Lytro} dataset consists of $100$ LF images for training and $30$ LF images for testing, which were captured using a \cite{Lytro} Illum provided by \cite{kalantari2016learning}. The \textit{HCI+Inria} dataset includes $22$ training LF images from the HCI LF dataset \cite{honauer2016dataset}, $33$ training LF images from the Inria Dense LF dataset \cite{shi2019framework}, $2$ testing LF images from the HCI LF dataset \cite{honauer2016dataset}, and $4$ testing LF images from the Inria Dense LF dataset \cite{honauer2016dataset}. Additionally, we evaluated our method on measurements captured by a real coded aperture camera \cite{inagaki2018learning} (details provided in Section \ref{sec:realcm}).

In LF denoising, we used the same settings as \cite{chen2018light} for the \textit{Stanford Archive} dataset, which includes $70$ LF images for training and $30$ LF images for testing. For the \textit{HCI+Inria} dataset, we used $55$ synthetic LF images of size $7 \times 7 \times 512 \times 512$ for training and $6$ synthetic LF images for testing.\\

\noindent \textbf{Training Settings.} 
During training, we calculated the $\mathcal{L}_1$ distance between $\mathbf{L^{(T)}}$  and the ground-truth LF as the loss function of our network.
In the training phase, we randomly cropped the training LF image into patches with size $M \times N \times 32 \times 32$.
The batch size was set to $5$. The training process consisted of $10K$ pre-training and $10K$ training epochs. We manually set all $\mathbf{U}$ and $\mathbf{V}$ to $1$ for pre-training. In the first $40\%$ training epoch, the value of $\tau$ decreased linearly from 1 to 0.05 and remained unchanged in the follow-up. We chose the Adam optimizer \cite{kingma2014adam} and used OneCycleLR \cite{smith2019super} to schedule the learning rate.
Our network was implemented with PyTorch.\\

\noindent \textbf{Network Settings.} We set $T ~~ =6$ and $J ~~ =8$ for CR-Net and $T ~~ =2$ and $J ~~ =6$ for our denoising network (see Sections \ref{sec:ablation1} and \ref{sec:ablation2} for the ablation study on the effect of $T$ and $J$). In each $\mathcal{D}^{(t)}(\cdot) ~~ ( 0 \leq t \leq T-1)$, the kernel size of four types of convolution methods was set to $3 \times 3$, and the output feature channel was set to $32$ and $64$ for CR-Net and DN-Net, respectively. Since $\mathcal{D}^{(t)}(\cdot)$ processes residual information, we removed all biases from the convolutional layers. It's worth noting that all projection layers $\mathcal{P}(\cdot)$ shared the same weights. Additionally, due to their physical interpretation, all parameters in $\mathcal{P}(\cdot)$ were clipped to the range of $[0,1]$ during the training process.

\begin{figure*}
\centering
\includegraphics[width=0.95\textwidth]{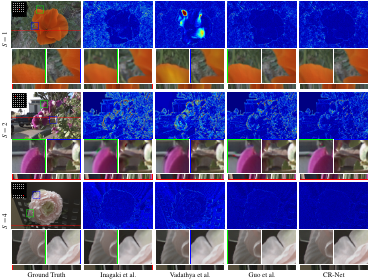}
\caption{Visual comparisons of reconstructed LFs from different methods under three tasks over the  \textit{Lytro} dataset. The error maps of SAIs and zoomed selected regions are shown for better visualization.}
\label{fig:IJCV_CA_lytrof1}
\end{figure*}

\begin{figure*}
    \centering
    \includegraphics[width=0.95\textwidth]{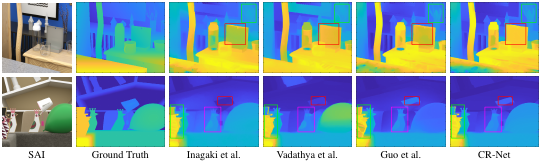}
    \caption{Visual comparisons of ground-truth depth maps and the estimated depth maps from the reconstructed LFs by different methods under task $S=2$ over the \textit{HCI+Inria} dataset.}
    \label{fig:IJCV_CA_depthf1}
\end{figure*}

\subsection{Evaluation on Compressive LF Reconstruction}
\subsubsection{Comparisons on Simulated CMs}
We compared our CR-Net with three state-of-the-art deep learning-based LF reconstruction methods from CMs, i.e., \cite{inagaki2018learning}, \cite{vadathya2019unified}, and \cite{guo2021deep}. For a fair comparison, all methods were re-trained on the same training datasets with officially released codes and recommended configurations. 
We conducted three tasks on the \textit{Lytro} dataset, i.e., reconstructing $7\times 7$ LFs with the input measurement number $S=1$, $S=2$, and $S=4$, respectively. We also conducted three tasks on the \textit{HCI+Inria} dataset, i.e., reconstructing $5\times5$ LFs with the input measurement number $S=1$, $2$, and $4$, respectively.\\

\noindent \textbf{Quantitative comparisons.} We separately calculated the average PSNR and SSIM over LFs in test datasets to quantitatively compare different methods. The results are shown in Table \ref{tab:table1}, where it can be observed that:
\begin{itemize}
    \item CR-Net achieves better performance with a smaller model size than all compared methods on most tasks, which gives credit to our physical interpretable cycle-consistent framework and the probabilistic-based feature embedding strategy;
    
    \item \cite{inagaki2018learning} has significantly lower PSNR and SSIM values than CR-Net on both \textit{Lytro} and \textit{HCI+Inria} datasets. The reason may be that \cite{inagaki2018learning} simply employs 2-D convolutional filters that are unable to model the 4-D LF well;
    
    \item \cite{vadathya2019unified} outperforms CR-Net under task $S = 1$ on the \textit{Lytro} dataset, which may be credited to its explicit use of geometry information. However, our CR-Net achieves much higher PSNR, i.e., about $5$ dB, than \cite{vadathya2019unified} under tasks $S = 2$ and $S = 4$ on the \textit{Lytro} dataset, because our probabilistic-based feature embedding strategy has the stronger representative ability to model the dimensional correlations among the LF with more than one CMs. Moreover, the proposed CR-Net significantly outperforms \cite{vadathya2019unified} under all tasks on the \textit{HCI+Inria} dataset. 
    We believe the reason is that  LFs contained in this dataset have relatively large disparity, resulting in heavily blurred CMs, and thus, \cite{vadathya2019unified} could have difficulties predicting a high-quality central SAI for view reconstruction; and 
    
    \item CR-Net outperforms \cite{guo2021deep} on both \textit{Lytro} and \textit{HCI+Inria} datasets. The possible reason is that \cite{guo2021deep} employs the SAS convolutional layer, which combines spatial and angular dimensions in an empirical manner, while our method can adaptively learn the fusion of the spatial-angular features.
    
\end{itemize}

\noindent \textbf{Qualitative comparisons.} Fig. \ref{fig:IJCV_CA_synf1} and Fig. \ref{fig:IJCV_CA_lytrof1} show visual comparisons of reconstructed LFs from different methods. We can observe that CR-Net produces better visual results than all the compared methods. Specifically, \cite{inagaki2018learning} and \cite{vadathya2019unified} show blurry effects, and lose high-frequency details at texture regions, while our CR-Net can reconstruct sharp details at both texture and smooth regions. Besides, \cite{guo2021deep} shows distortions at occlusion boundaries, while our CR-Net can reconstruct clearer and more accurate structures.\\

\begin{figure*}[t]
    \centering
    \includegraphics[width=0.8\linewidth]{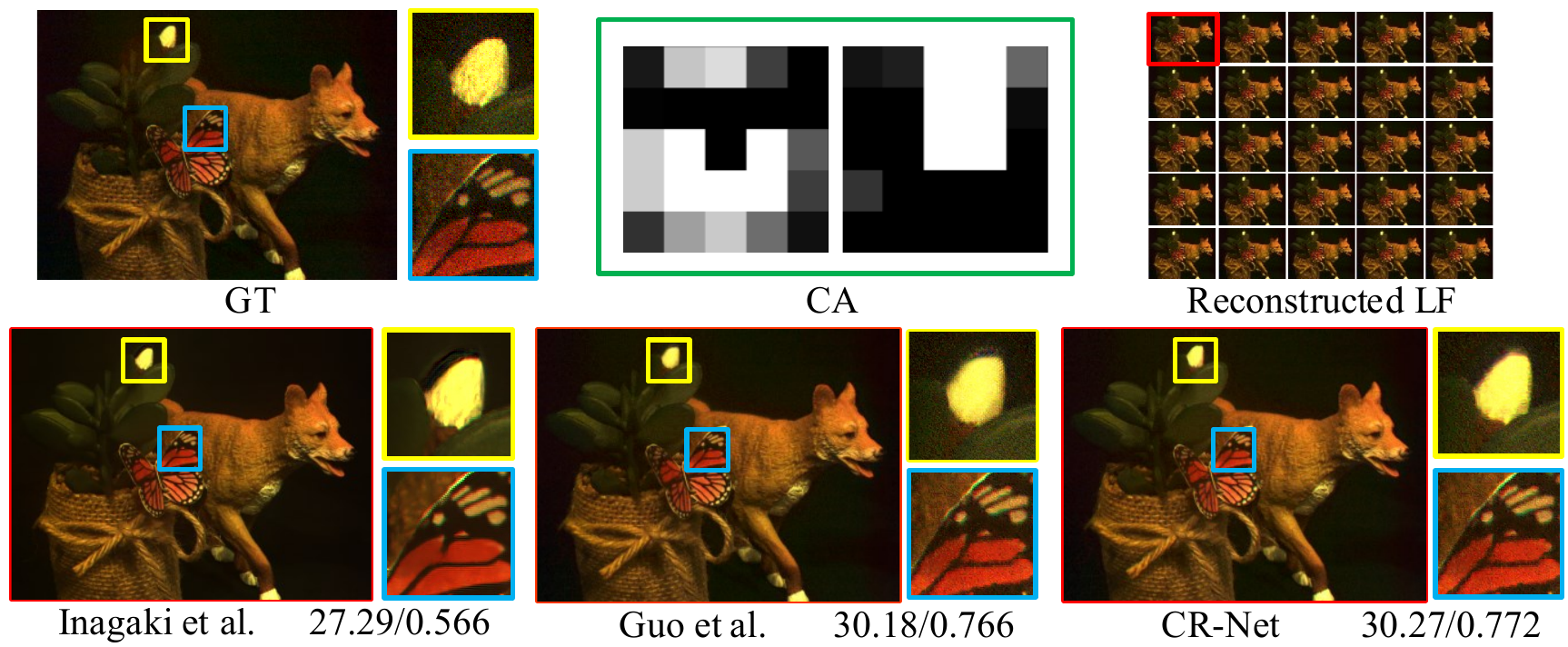}
    \caption{Comparisons of our CR-Net against \cite{inagaki2018learning} and \cite{guo2021deep} on real CMs. The PSNR/SSIM values are calculated between the reconstructed LF and the pinhole images.}
    \label{fig:real-ca}
\end{figure*}

\noindent \textbf{Comparisons of the LF parallax structure.} 
The quality of the parallax structure is one of the most important criteria for evaluating the reconstructed LF. To compare the ability to preserve the LF parallax structure, we visualized EPIs of the reconstructed LFs using different methods. As shown in Fig. \ref{fig:IJCV_CA_synf1} and Fig. \ref{fig:IJCV_CA_lytrof1}, we observe that the proposed CR-Net shows clearer and sharper linear structures than the compared methods, which demonstrates better preservation of the parallax structures of reconstructed LFs. Moreover, the accuracy of the depth map estimated from the reconstructed LF reflects how well the parallax structure is preserved to some extent. Therefore, we compared the depth maps estimated from different methods using the same depth estimation method \cite{chen2018accurate} and the ground truth depth maps. As shown in Fig. \ref{fig:IJCV_CA_depthf1}, we can see that the depth maps of CR-Net are more similar to those of the ground truth at both smooth regions and occlusion boundaries. This means that our method can better preserve the parallax structure than other methods.

\subsubsection{Comparisons on Real CMs}
\label{sec:realcm}
To demonstrate the ability of our method on real CMs, we adopted the real data captured by a coded aperture camera built by Inagaki \textit{et al.} \cite{inagaki2018learning}.
The real data contains $2$ CMs, $2$ masks that are used for capturing the CMs, and the captured $5 \times 5$ pinhole images as the ground truth. We fixed the weights of $\mathcal{P}$ to the mask values, and
re-trained the CR-Net under the task $S=2$ on the \textit{HCI+Inria} dataset.
Then, we tested the trained model with the real CMs, and compared the reconstructed LF by our CR-Net against that of \cite{inagaki2018learning} and \cite{guo2021deep}. The results are shown in Fig. \ref{fig:real-ca}, where it can be observed that the reconstructed LF from our CR-Net is closer to the ground-truth one, while the results from \cite{inagaki2018learning} show blurring artifacts at the occlusion boundaries, demonstrating the advantage of our CR-Net on real CMs. 

\begin{figure}[t]
    \centering
\includegraphics[width=0.95\linewidth]{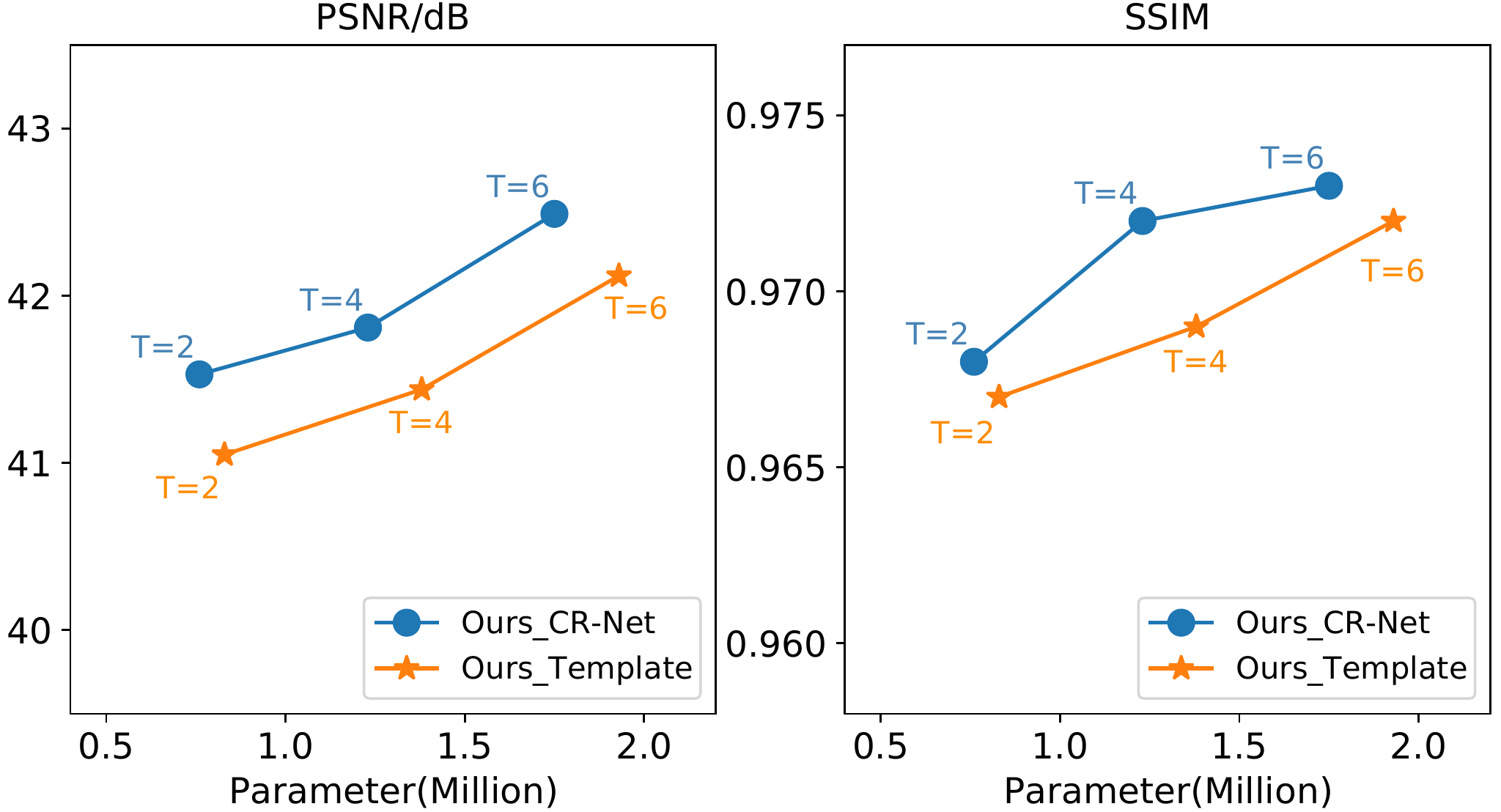}
    \caption{
    Quantitative results of our final network, denoted as Ours\_CR-Net, and the template network without probabilistic modeling, denoted as Ours\_Template, under different numbers of stages. The PSNR/SSIM values under the task $S=2$ on the \textit{HCI+Inria} dataset as well as the model size are provided for comparisons.
    }
    \label{fig:result4}
\end{figure}

\subsubsection{Ablation Study}
\label{sec:ablation1}

\textbf{The number of stages and feature embedding units for the CR-Net.} 
To investigate the influence of the number of stages and feature embedding units on the performance, we compared the results of CR-Net using $T=2$, $4$ and $6$ and $J=6$, $8$, and $10$. As demonstrated in Fig. \ref{fig:result4} and Table \ref{tab:CA-1}, the PSNR/SSIM values gradually improve from $2$ to $6$ stages for our CR-Net and the template network. Moreover, $8$ feature embedding units yield better reconstruction quality. Consequently, for superior LF reconstruction quality, we selected $6$ stages and $8$ feature embedding units in our CR-Net.

\begin{figure}[t]
    \centering
    \includegraphics[width=\linewidth]{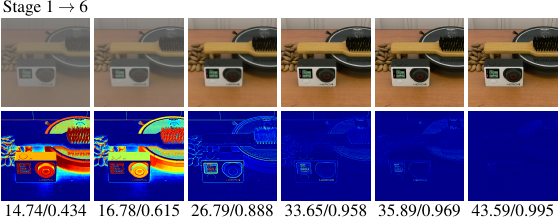}
    \caption{Visualization of the intermediate output at each stage of CR-Net. The first row presents central SAIs of the output LFs, and the second row presents error maps between the reconstructed central SAIs and ground-truth ones. The PSNR/SSIM values of the output LFs at different stages are listed below.}
    \label{fig:stage-num}
\end{figure}

Additionally, to further demonstrate the effectiveness of the iterative refinement framework, we visualized the output of each stage of our CR-Net.
As shown in Fig. \ref{fig:stage-num}, the quality of the reconstructed LF gradually improves from the first stage to the last one, demonstrating the effectiveness of the proposed progressive refinement framework based on the cycle-consistency.\\

\begin{table}
\centering
  \caption{
    Performance comparisons of our CR-Net with different numbers of stages and feature embedding units in a probabilistic-based feature embedding module.
  }
  \label{tab:CA-1}
  \begin{tabular}{cc|cc|c}
    \toprule[1.2pt]
     & & PSNR & SSIM & \# Parameters \\ \hline
    % \midrule
    \multirow{3}*{$T = 6$} 
     & $J = 6$  & 41.55 & 0.968 & 1.41\\
     & $J = 8$  & 42.49 & 0.973 & 1.75\\
     & $J = 10$ & 41.91 & 0.969 & 2.48\\ \hline
    \multirow{3}*{$J = 8$} 
     & $T = 2$  & 41.53 & 0.968 & 0.82\\
     & $T = 4$  & 41.81 & 0.972 & 1.38\\
     & $T = 6$  & 42.49 & 0.973 & 1.75\\ \bottomrule[1.2pt]
\end{tabular}
\end{table}

\begin{figure}
  \centering
  \includegraphics[width=0.95\linewidth]{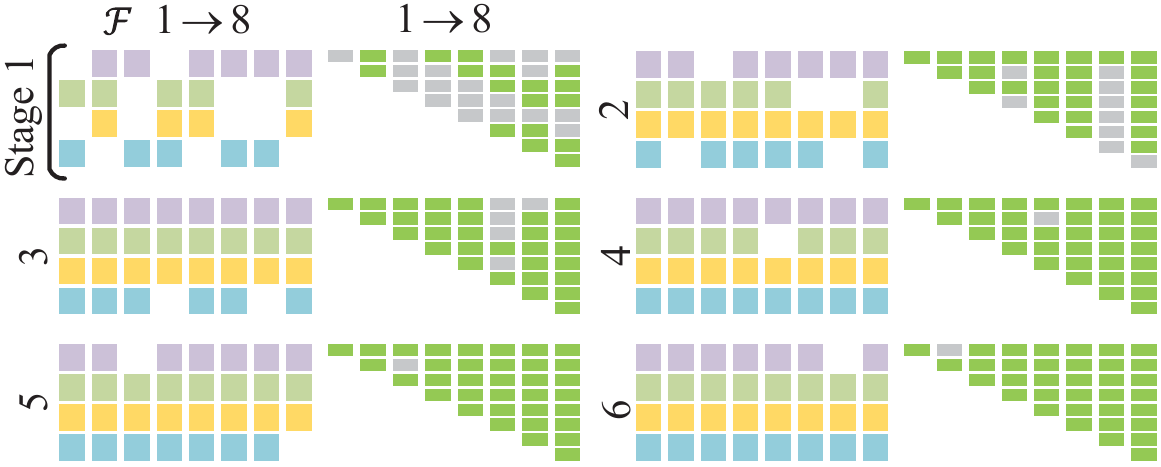}
  \caption{Visualization of the learned architecture for the probabilistic-based feature embedding in the CR-Net for task $S=2$ over the \textit{HCI+Inria} dataset. The CR-Net contains $6$ stages and $8$ feature embedding units in each stage. 
  For each stage, we show the reserved convolution patterns by patches with different colors on the left, and the network-level feature aggregation results on the right accordingly, where the used features are marked in green.}
  \label{fig:plot-network}
\end{figure}

\noindent \textbf{Effectiveness of modeling probabilistic space.} 
In Fig.~\ref{fig:plot-network}, we visualized the finally learned architecture for the probabilistic-based feature embedding module, where it can be seen that both the layer-wise feature extraction and network-level feature aggregation are adaptively constructed based on the probabilistic modeling.
Furthermore, to evaluate the effectiveness of modeling probabilistic space, we quantitatively compared the performance of the final learned CR-Net against the template network.
As shown in Fig.~\ref{fig:result4} and Table \ref{tab:table4}, it can be observed that the final CR-Net consistently achieves higher performance with a smaller model size compared with the template network under different numbers of stages, which demonstrates the effectiveness of modeling probabilistic space for space-angular fusion in our method.\\

\noindent \textbf{Effectiveness of the probabilistic-based feature embedding module.} 
To verify the effectiveness of the probabilistic-based feature embedding module $\mathcal{D}(\cdot)$, we replaced $\mathcal{D}(\cdot)$ in each stage with either a stack of 3-D convolution layers or a stack of SAS layers while keeping a comparable model size to our CR-Net.
The resulting networks are denoted as CR-Conv3D and CR-SAS, respectively, and their performance was compared with our CR-Net in Table \ref{tab:table4}. The results indicate that replacing the probabilistic-based feature embedding module with 3-D convolution or SAS layers leads to obvious performance degradation, which further demonstrates the advantage of the proposed probabilistic-based feature embedding strategy.\\ 

\noindent \textbf{Effectiveness of sharing weights in projection layers.} 
We also verified the effectiveness of sharing weights of the projection layers across different stages. As shown in Table \ref{tab:table4}, the performance of the model without sharing the weights of the projection layers, denoted as w/o sharing $\mathcal{P}(\cdot)$, is lower than the CR-Net. 
The reason is that sharing weights guarantees that the projection layers across all stages correspond to the same physical imaging process. As the pseudo-CMs are produced under the same projection, the residual between the pseudo-CMs and the input CMs is consistent and can be minimized progressively.

\begin{table}
\centering
  \caption{
    Effectiveness of the probabilistic-based feature embedding module and sharing the weights of projection layers in the proposed CR-Net. 
    The results on task $S=2$ over the \textit{HCI+Inria} dataset are provided for comparisons.
  }
  \label{tab:table4}
  \begin{tabular}{c|cc|c}
    \toprule[1.2pt]
     & PSNR & SSIM & \# Parameters \\
    \hline
    w/o sharing $\mathcal{P}(\cdot)$ & 41.67 & 0.970 & 1.75\\
    CR-Conv3D(1)  & 40.42  & 0.963 & 1.78 \\
    CR-Conv3D(2)  & 41.34  & 0.966 & 2.67 \\
    CR-SAS       & 41.11 & 0.966 & 1.78\\
    CR-Template & 42.12 & 0.972 & 1.83\\
    CR-Net & \textbf{42.49} & \textbf{0.973} & 1.75 \\
  \bottomrule[1.2pt]
\end{tabular}
\end{table}

\begin{table*}
\centering
\caption{Quantitative comparisons (PSNR/SSIM) of different methods over the test LFs in \textit{Stanford Archive} and \textit{HCI+Inria} datasets under the denoising task of three noise levels. The number of network parameters (M) of different methods is also provided. “-” indicates that
the method is not a deep learning-based method.}
\label{tab:dn-all}
  \begin{tabular}{c|c|ccc|c}
    \toprule[1.2pt]
    Dataset & Method & $\sigma = 10$  & $\sigma = 20$ & $\sigma = 50$ & \# Parameters \\\hline
    \multirow{7}{*}{\textit{Stanford Archive}} 
    & DeamNet \cite{ren2021adaptive} & 37.22/0.971 & 33.69/0.941 & 29.38/0.876 & 2.23\\
    & LFBM5D \cite{alain2017light} & 38.25/0.927 & 33.27/0.841 & 25.11/0.642 & - \\
    & Mviden \cite{tomita2022denoising} & 39.40/0.916 & 36.67/0.861 & 32.92/0.742 & - \\
    & APA \cite{chen2018light} & 39.34/0.956 & 36.94/0.939 & 33.65/0.894 & 1.25\\
    & DSAR \cite{guo2021deep} & 42.77/0.973 & 39.98/0.958  & 35.96/0.920 & 2.22\\
    & MSP \cite{wang2023multi} & 42.37/0.974 & 39.67/0.962  & 36.06/0.929 & 1.25\\
    & DN-Net & \textbf{43.20}/\textbf{0.978} & \textbf{40.59}/\textbf{0.964} & \textbf{36.72}/\textbf{0.931} & 2.39 \\ \hline
    \multirow{7}{*}{\textit{HCI+Inria}} 
    & DeamNet \cite{ren2021adaptive} & 38.05/0.976 & 34.71/0.955 & 30.65/0.919  & 2.23\\
    & LFBM5D \cite{alain2017light} & 40.83/0.965 &  37.36/0.936 & 31.94/0.860 & - \\
    & Mviden \cite{tomita2022denoising} & 38.95/0.940 & 35.93/0.907 & 31.96/0.831 & - \\
    & APA \cite{chen2018light} & 36.89/0.933 & 33.92/0.896 & 30.19/0.813 & 1.22 \\
    & DSAR \cite{guo2021deep} & 42.10/0.971 & 39.53/0.955 & 35.14/0.899 & 2.22\\
    & MSP \cite{wang2023multi} & 41.51/0.972 & 38.80/0.956 & 35.31/0.913 & 1.23\\
    & DN-Net & \textbf{43.21}/\textbf{0.976} & \textbf{40.50}/\textbf{0.962} & \textbf{36.24}/\textbf{0.918} & 2.39 \\
  \bottomrule[1.2pt]
\end{tabular}
\end{table*}

\subsection{Evaluation on 4-D LF Denoising}

We compared our DN-Net against several state-of-the-art methods, including a local similarity-based LF denoising method denoted by LFBM5D \cite{alain2017light}, a short-time DFT approach denoted by Mviden \cite{tomita2022denoising}, two deep learning-based LF denoising methods named APA \cite{chen2018light} and MSP \cite{wang2023multi}, a deep unrolling-based LF denoising method DSAR \cite{guo2021deep}, and a deep learning-based single image denoising method DeamNet \cite{ren2021adaptive} which was applied on each SAI of the input LF as a baseline.

The same noise synthesis and preprocessing protocol as APA were used in our experiment, i.e., adding zero-mean Gaussian noise with the standard variance $\sigma$ varying in the range of 10, 20, and 50 to generate the noisy LF images. We use the same training and test datasets with a narrow baseline from Stanford Lytro Light Field Archive \cite{StanfordArchive} as APA, extracting central $8 \times 8$ SAIs and converting them to grayscale for each LF. In addition, we performed experimental validation on the wider baseline \textit{HCI+Inria} dataset, in which central $7 \times 7$ SAIs were used, $55$ and $6$ LF images were used for training and testing, respectively. 
We trained all the comparison methods for each noise level, and evaluated each model with the matched noise level on test data.

\begin{table}
\caption{Comparisons of the running time (in seconds) of different methods for denoising tasks on \textit{HCI+Inria}.}
\centering
\label{tab:my-time}
\begin{tabular}{cc|c} \toprule[1.2pt]
\multicolumn{2}{c|}{Method} & Running Time \\ \hline  
DeamNet         & \cite{ren2021adaptive}   &  2.35 \\ 
LFBM5D          & \cite{alain2017light}    &  79.65 \\
Mviden          & \cite{tomita2022denoising}    &  118.51 \\
APA             & \cite{chen2018light}    &  17.11 \\
DSAR            & \cite{guo2021deep}    &  14.72 \\
MSP             & \cite{wang2023multi}    &  4.16 \\
\multicolumn{2}{c|}{Proposed DN-Net}    &  11.63 \\
\bottomrule[1.2pt]
\end{tabular}
\end{table}

\subsubsection{Quantitative Comparison}
\begin{figure}
  \centering
  \includegraphics[width=0.95\linewidth]{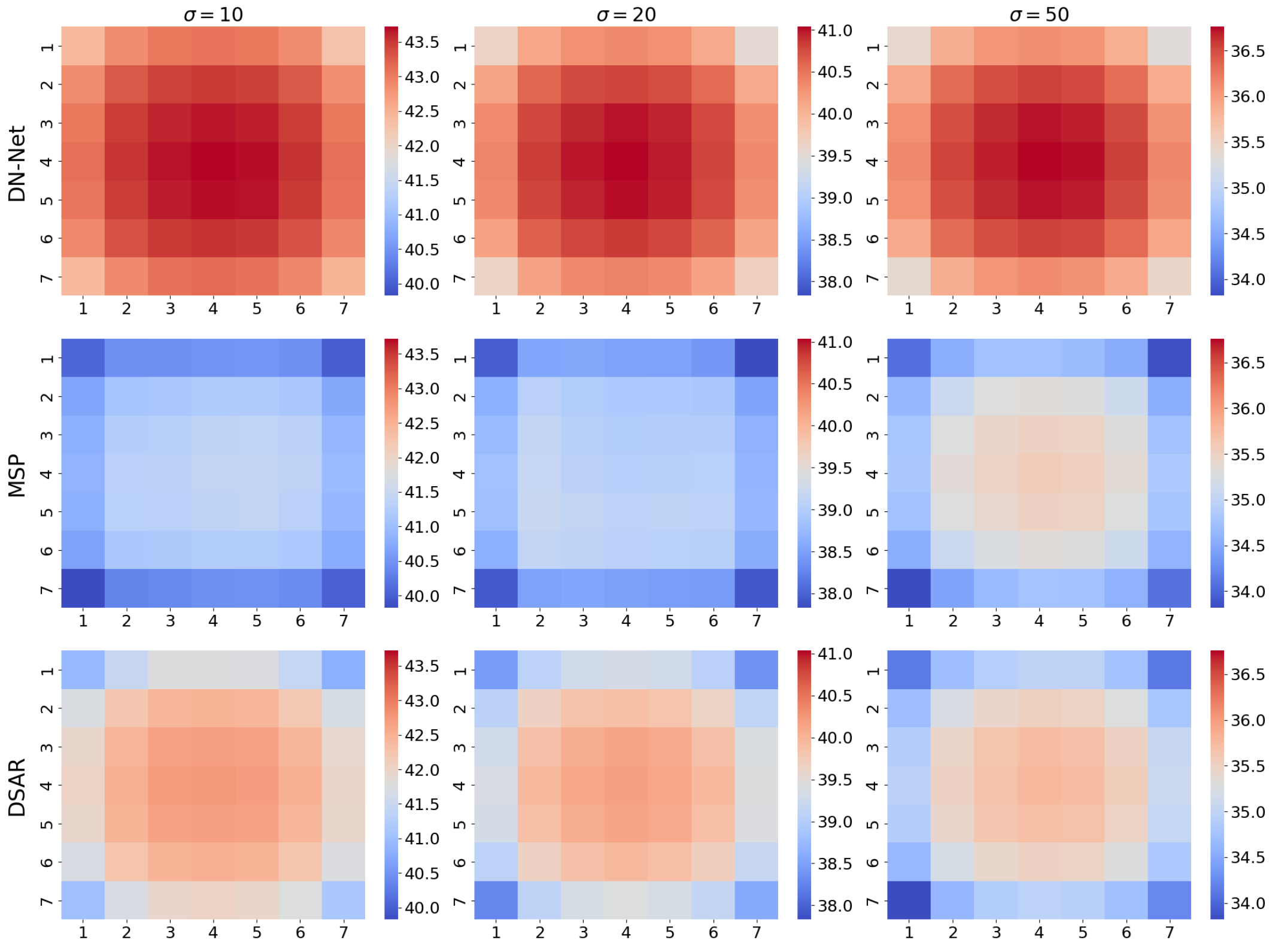}
  \caption{Comparison of the average PSNR for each SAI of denoised LFs produced by various methods. The PSNR values depicted are calculated as the average across the test set derived from the \textit{HCI+Inria} dataset.
  }
  \label{fig:dn-sai}
\end{figure}

The average PSNR and SSIM between the denoised LFs and ground-truth ones were used to evaluate different methods quantitatively. From Table \ref{tab:dn-all}, it can be observed that the proposed DN-Net outperforms the second-best method, i.e., DSAR, up to $0.5dB$ on \textit{Stanford Archive} for all three noise levels, and up to $0.9dB$ on \textit{HCI+Inria} for all three noise levels, validating the advantage of our network. Besides, the performance advantage of DN-Net is more obvious on wider baseline LFs, validating our method's ability to process large baseline datasets. Additionally, Fig. \ref{fig:dn-sai} depicts the average PSNR at each SAI position of the denoised LFs by different methods. The performance of our DN-Net across different SAIs exhibits greater consistency and stability when compared to other methods. This demonstrates the robustness of our approach in maintaining high PSNR values for varying SAIs.

\begin{figure*}
    \centering
    \includegraphics[width=1\textwidth]{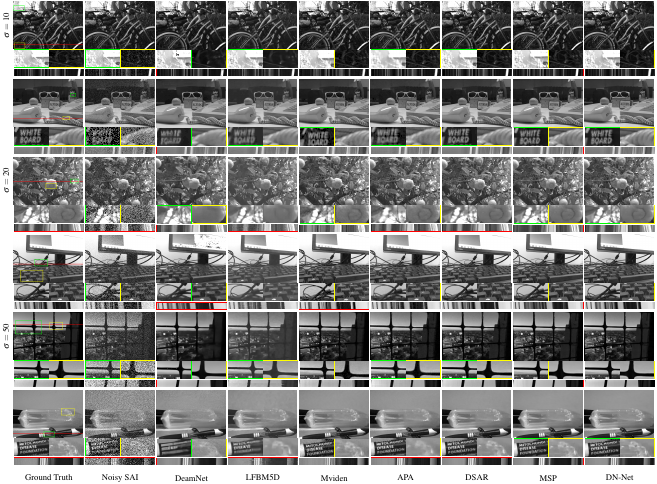}
    \caption{Visual comparisons of denoised LF from different methods over the \textit{Stanford Archive} dataset. The zoomed selected regions and EPI are shown for better visualization}
    \label{fig:dn-lytro}
\end{figure*}

\begin{figure*}
    \centering
    \includegraphics[width=1\textwidth]{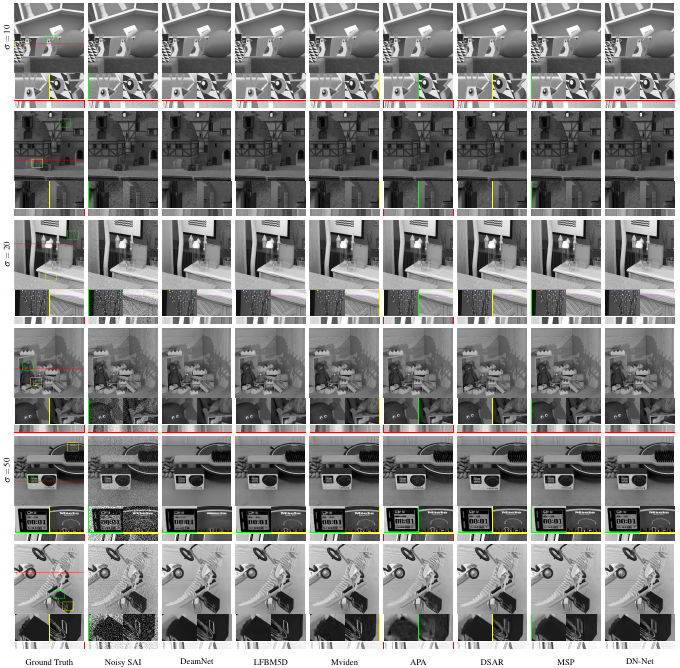}
    \caption{Visual comparisons of denoised LF from different methods over the \textit{HCI+Inria} dataset. The zoomed selected regions and EPI are shown for better visualization}
    \label{fig:dn-syn}
\end{figure*}

\begin{figure*}
    \centering
    \includegraphics[width=1\textwidth]{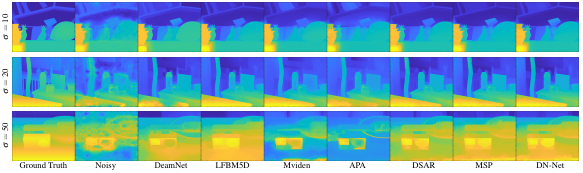}
    \caption{Visual comparisons of ground-truth depth maps and estimated depth maps from denoised light fields by various methods on the \textit{HCI+Inria} dataset.}
    \label{fig:dn-syn-depth}
\end{figure*}

\subsubsection{Visual Comparison}
Fig. \ref{fig:dn-lytro} and Fig. \ref{fig:dn-syn} present visual comparisons of denoised LFs obtained through different methods across three noise levels $\sigma = 10, 20$ and $50$. The results indicate that DeamNet, LFBM5D, and Mviden fail to preserve high-frequency details, such as the texture regions of flowers, and generate severe distortions, such as the window frames. Additionally, DeamNet produces hole artifacts in the denoised LFs. While APA yields comparatively better visual results than LFBM5D and DeamNet, it still exhibits blurring artifacts at occlusion boundaries. DSAR and MSP achieve relatively good visual results, but the details in textured regions and occlusion boundaries are not as sharp as those produced by our DN-Net. Visual comparison of the EPIs extracted from the denoised LFs indicates that our method can preserve sharper linear structures, thereby validating its advantage in preserving LF parallax structures during denoising.

Furthermore, we compared the estimated depth maps obtained through different methods using the same depth estimation method \cite{chen2018accurate}. As shown in Fig. \ref{fig:dn-syn-depth}, the depth maps generated by our DN-Net are more similar to the ground truth at both smooth regions and occlusion boundaries, indicating that the proposed DN-Net can better preserve the parallax structure than the other methods.

\begin{figure*}
    \centering
    \includegraphics[width=0.95\textwidth]{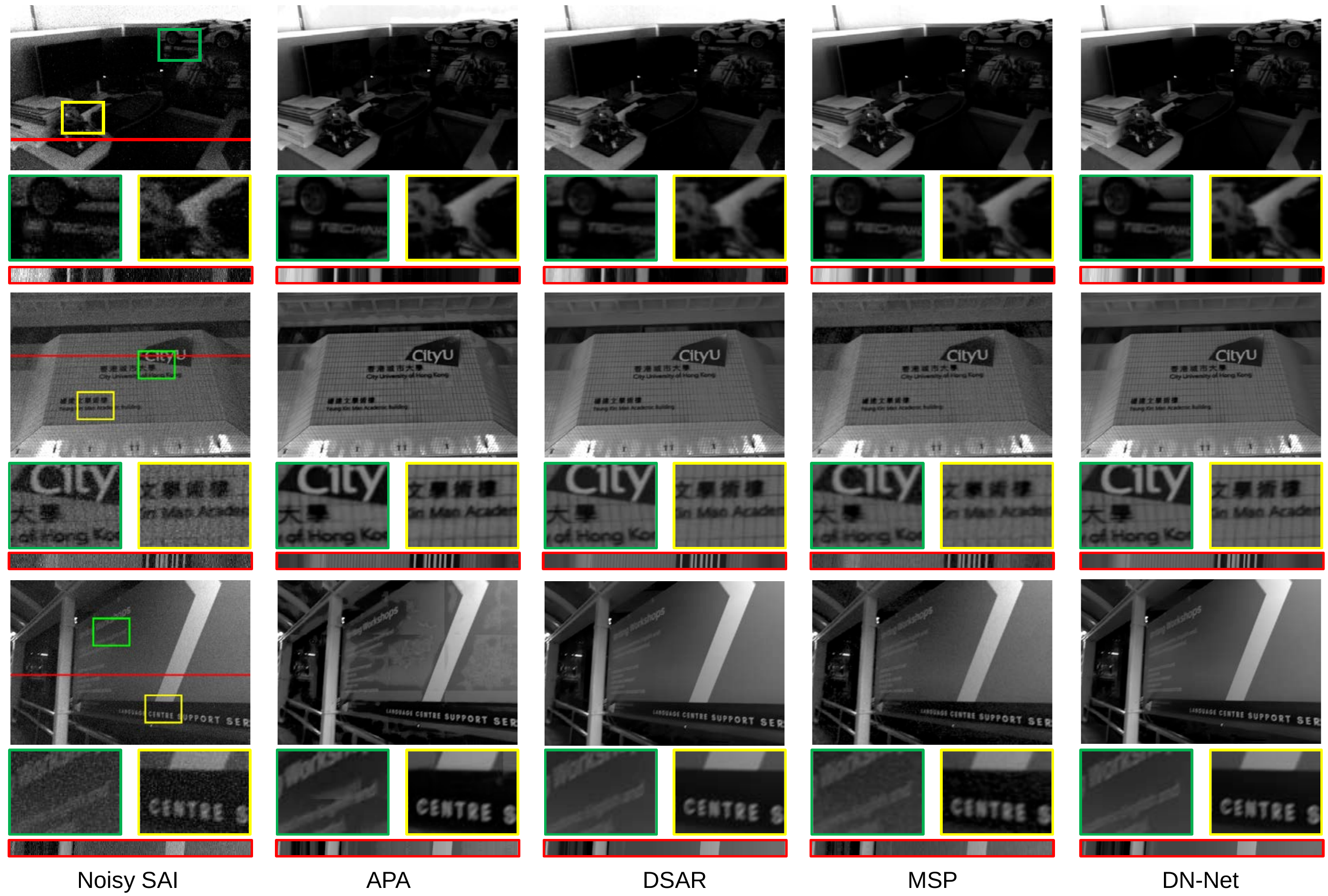}
    \caption{Visual comparisons of different methods on real LF denoising. Our DN-Net trained on \textit{Stanford Archive} dataset with $\sigma = 20$ were selected.
    The zoomed selected regions and EPI are shown for better visualization}
    \label{fig:dn-illum}
\end{figure*}

\subsubsection{Comparisons on Real LF Denoising}
When the lighting of the shooting scene is not sufficient, the light field data will be seriously affected by noise.  Therefore, we used the low-light data collected by the Lytro Illum camera and verified the effectiveness of our DN-Net trained on the \textit{Stanford Archive} dataset and with $\sigma = 20$. From the experimental results in Fig. \ref{fig:dn-illum}, we can observe that our algorithm can effectively remove noise while maintaining clear scene details better.

\subsubsection{Comparison of Running Time}
We compared the inference time (in seconds) of different methods for LF denoising under \textit{HCI+Inria} dataset, and Table \ref{tab:my-time} lists the results. All methods were tested on a desktop with Intel CPU i7-8700 @ 3.70GHz, 32 GB RAM and NVIDIA GeForce RTX 3090. As shown in Table \ref{tab:my-time}, Ours is faster than \cite{alain2017light}, \cite{tomita2022denoising}, \cite{chen2018light} and \cite{guo2021deep} but slower than \cite{ren2021adaptive} and \cite{wang2023multi}.

\subsubsection{Ablation Study}
\label{sec:ablation2}
We conducted ablation studies on the \textit{HCI+Inria} dataset with $\sigma = 20$ to verify the effectiveness of the extended framework and the impact of the number of stages and feature embedding units.\\

\noindent\textbf{The number of stages and feature embedding units.} We varied the number of iterative stages in the range from 1 to 4 with 6 fixed feature embedding units, and varied the number of feature embedding units, i.e., 4, 6, and 8 with 2 fixed iterative stages. As shown in Table \ref{tab:dn-1}, the denoising performance improves with the increase in the number of iterative stages. Moreover, the improvement is more obvious from 1 stage to 3 stages, while slight from 3 stages to 4 stages. What's more, the performance improves with the number of feature embedding units increasing. To balance the denoising performance and the number of network parameters, we finally choose 2 stages with 6 feature embedding units to process all the LF denoising tasks.

The application of the proposed probabilistic-based feature embedding module to various LF tasks requires a balance between performance and computational efficiency. As a general guideline, we recommend maintaining the number of network parameters within the range of $2M \sim 4M$. This can be achieved using configurations such as $T = 3, J = 6, C = 64$ or $T = 6, J = 6, C = 32$, where $C$ represents the number of feature channels. These initial configurations can be fine-tuned further as per specific requirements.\\ 

\begin{table}
\centering
  \caption{
    Performance comparisons of our LF denoising method with different numbers of iterative stages and feature embedding units in a probabilistic-based feature embedding module.
  }
  \label{tab:dn-1}
  \begin{tabular}{cc|cc|c}
    \toprule[1.2pt]
     & & PSNR & SSIM & \# Parameters \\ \hline
    % \midrule
    \multirow{3}*{$T$ = 2} 
     & $J$ = 4  & 40.15 & 0.959 & 1.57\\
     & $J$ = 6  & 40.50 & 0.962 & 2.39\\
     & $J$ = 8  & 40.65 & 0.963 & 3.32\\ \hline
    \multirow{4}*{$J$ = 6} 
     & $T$ = 1  & 40.01 & 0.958 & 1.61\\
     & $T$ = 2  & 40.50 & 0.962 & 2.39\\
     & $T$ = 3  & 40.72 & 0.963 & 3.23\\
     & $T$ = 4  & 40.76 & 0.963 & 4.03\\ \bottomrule[1.2pt]
\end{tabular}
\end{table}

\noindent\textbf{Effectiveness of the proposed noise suppression module and probabilistic-based feature embedding.} 
Our approach combines a noise suppression module with the proposed probabilistic-based feature embedding for LF denoising. To validate the effectiveness of the proposed noise suppression module, we conducted a quantitative comparison of the quality of denoised LF images generated by our method without the noise suppression module (i.e., w/o Noise Suppression). 

Furthermore, to evaluate the effectiveness of the proposed probabilistic-based feature embedding, we replaced the probabilistic-based feature embedding $\mathcal{D}(\cdot)$ in each stage with either a stack of 3-D convolution layers or SAS layers while maintaining a similar model size to our framework. The resulting networks were denoted as DN-Conv3D and DN-SAS, respectively, and their performances were compared with the proposed DN-Net in Table \ref{tab:dn-2}. 

\begin{figure}
    \centering
    \includegraphics[width=0.95\linewidth]{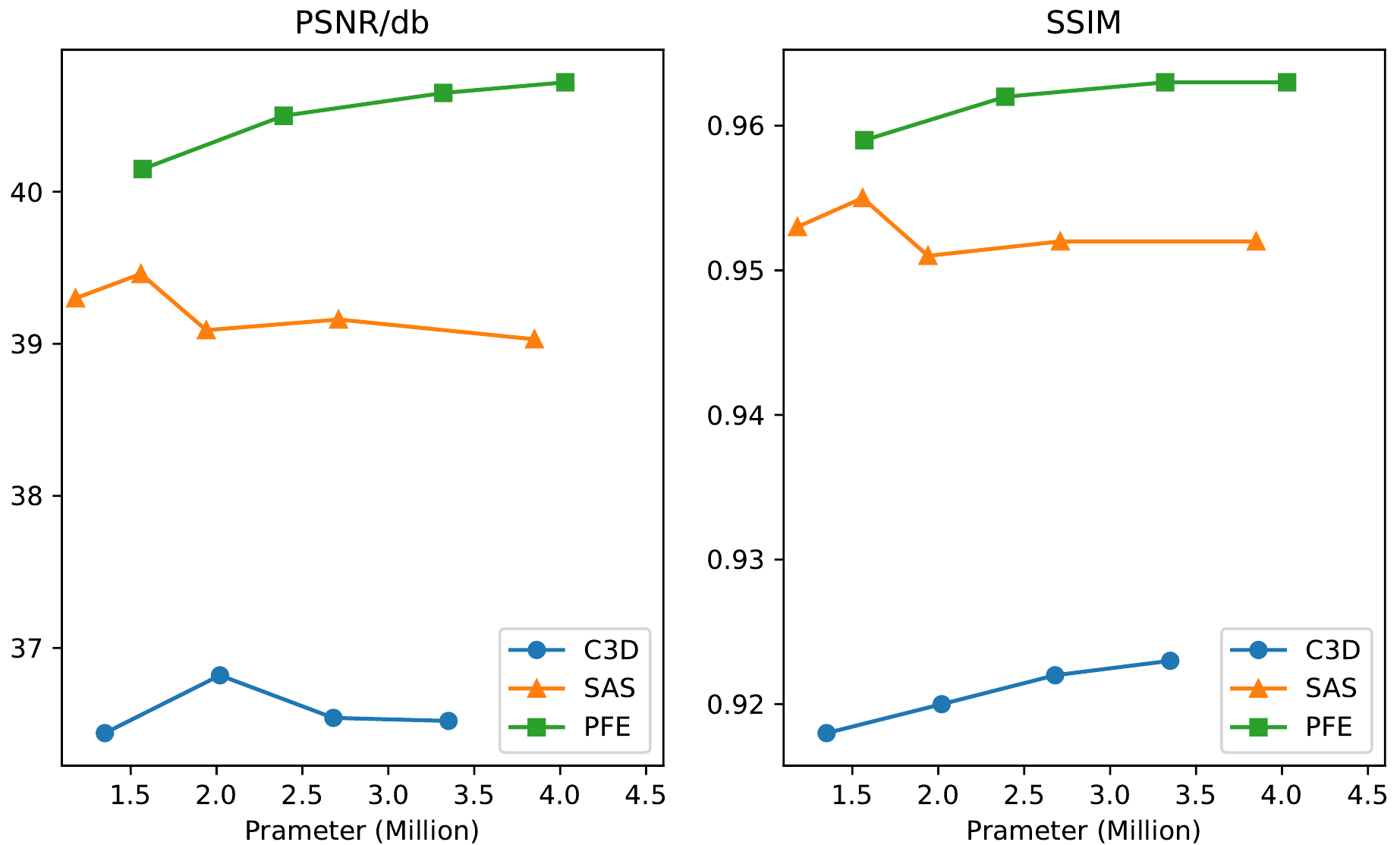}
    \caption{
    Quantitative results of the proposed denoising framework equipped with 
    different feature embedding extraction methods, i.e., the proposed probabilistic-based feature embedding (PFE), the SAS convolution (SAS), and the 3-D convolution (C3D).
    The PSNR/SSIM values for the task $\sigma =20$ on the \textit{HCI+Inria} dataset, as well as the model size, are provided for comparisons.
    }
    \label{fig:dn-line}
\end{figure}

\begin{table*}[t]
\centering
  \caption{
    Comparative analysis of the effectiveness of the proposed PFE module on LF spatial and angular super-resolution (SR). We selected $2\times$ spatial SR with $5\times5$ SAIs and $2\times2 \rightarrow 7\times7$ angular SR for the comparison. The table presents the PSNR/SSIM results for two tasks, DistgSSR and DistgASR, with and without the integration of PFE.
  }
  \label{tab:lfsr}
  \begin{tabular}{cc|ccccc}
    \toprule[1.2pt]
    $2\times$ & Model  & EPLF & HCInew & HCIold & Inria & STFgtr \\ \hline
    \multirow{2}*{Spatial SR} 
    & DistgSSR & 34.58/0.977 & 37.76/0.979  & 44.77/0.995 & 36.48/0.985  & 40.29/0.994    \\
    & DistgSSR-PFE & 34.85/0.978 & 37.92/0.979  & 44.86/0.995 & 36.69/0.986  & 40.55/0.994    \\
    \hline
    $2\times2 \rightarrow 7\times7$ & Model  & HCInew & HCIold & 30scenes & Occlusion & Reflective  \\ \hline
    \multirow{2}*{Angular SR} 
    & DistgASR  & 34.28/0.970  & 42.12/0.977  & 43.36/0.995  & 39.15/0.990  & 38.85/0.977\\
    & DistgASR-PFE  & 34.52/0.972  & 42.33/0.979  & 43.64/0.995  & 39.40/0.991  & 39.08/0.978    \\
  \bottomrule[1.2pt]
\end{tabular}
\end{table*}

Moreover, we proposed a baseline framework without the proposed modules and replaced the feature embedding units with SAS. As shown in Table \ref{tab:dn-2}, it can be observed that without the proposed noise suppression module, the value of PSNR decreased by approximately $0.45 dB$, thereby validating its effectiveness. Furthermore, the proposed DN-Net outperformed the 3-D convolution, SAS, and template network (DN-Template) in our evaluations, demonstrating the advantage of the proposed probabilistic-based feature embedding. Moreover, the effectiveness of the overall framework was validated by comparing the baseline with the proposed DN-Net, where it can be seen that the value of PSNR improved by $2.3 dB$.

To gain further insight into the advantages and disadvantages of commonly used LF feature extraction methods, i.e., the proposed probabilistic-based feature embedding (PFE), SAS convolution (SAS), and 3-D convolution (C3D), we present a performance analysis with parameter quantities in Fig. \ref{fig:dn-line}.
As demonstrated, the performance of our proposed PFE steadily improves as the number of parameters increases. Whereas the PSNR of SAS and C3D initially improves, but then the performance fluctuates or declines, possibly due to network overfitting. Therefore, our proposed PFE is more suitable for building deeper networks.

\begin{table}
\centering
  \caption{
    Effectiveness of the proposed framework in terms of average PSNR and SSIM values.
  }
  \label{tab:dn-2}
  \begin{tabular}{c|cc|c}
    \toprule[1.2pt]
     & PSNR & SSIM & \# Parameters\\
    \hline
    w/o Noise Suppression & 40.04 & 0.959 & 2.38\\
    Baseline & 38.14 & 0.942 & 2.46\\
    DN-Conv3D(1) & 36.82 & 0.920 & 2.02\\
    DN-Conv3D(2) & 36.54 & 0.922 & 2.68\\
    DN-SAS & 38.82 & 0.950 & 2.32\\
    DN-Template & 40.27 & 0.960 & 2.41\\
    DN-Net & \textbf{40.50} & \textbf{0.962} & 2.39 \\
  \bottomrule[1.2pt]
\end{tabular}
\end{table}

\subsection{Evaluation on 4-D LF Super-Resolution}
In addition to the coded aperture-based compressive LF reconstruction, LF super-resolution approaches, including spatial and angular super-resolution techniques, present an alternative route to achieving high-quality LFs. The spatial-angular trade-offs present in LF cameras have incited the development of several LF super-resolution methods to enhance spatial or angular resolution.

To further validate the efficacy of our Probabilistic-based Feature Embedding (PFE) approach, we conducted experiments based on a recent work \cite{wang2022disentangling}. We specifically applied PFE to two distinct tasks, namely DistgSSR and DistgASR, and retrained the models using the dataset and codes officially released. The results, which provide a performance baseline, are documented in Table \ref{tab:lfsr}.
To control for potential confounding effects of different network settings (e.g., the use of LeakyReLU), we restructured DistgSSR and DistgASR into template networks and integrated our PFE methodology, yielding DistgSSR-PFE and DistgASR-PFE. In the detailed implementation, PFE is deployed on the Disentangling block of the DistgSSR-PFE network and on both the Disentangling block and Disentangling group of the DistgASR-PFE network. The experimental findings, as outlined in Table \ref{tab:lfsr}, clearly demonstrate that our PFE can enhance network performance.

\section{Conclusion} \label{sec7}
We have presented a novel adaptive probabilistic-based feature embedding for LFs. To verify its effectiveness, we evaluated its performance on compressive LF reconstruction and LF denoising, respectively.
For compressive LF imaging, we proposed a physically interpretable cycle-consistent framework that incorporates probabilistic-based feature embedding.
Our approach can reconstruct 4-D LFs from 2-D measurements with higher quality, improving the PSNR value by approximately 4.5 dB while preserving the LF parallax structure better than state-of-the-art methods.
Furthermore, we demonstrate the efficacy of our probabilistic-based feature embedding approach for LF denoising through a carefully designed network. Extensive experiments on diverse datasets demonstrate that our method outperforms state-of-the-art approaches by up to $1.1$ dB.

In future work, the application of our proposed probabilistic-based feature embedding could be potentially broadened to include other tasks within the domain of LF image processing, such as depth estimation.
Moreover, our proposed CR-Net exhibits significant potential for direct integration in reconstruction-based LF compression tasks \cite{hou2018light}. The inherent capabilities of CR-Net could be leveraged to gather, compress, transmit, and subsequently reconstruct compressed sensing images. This process would yield high-quality light fields, all while necessitating less data.\\

\noindent\textbf{Data Availability Statements}

As indicated in the manuscript, the used datasets are deposited in publicly available repositories.\\

\noindent \textbf{Conflict of Interest}

The authors declare that they do not have any commercial
or associative interest that represents a conflict of interest in connection with the work submitted.

\bibliographystyle{unsrtnat}      % basic style, author-year citations
\bibliographystyle{spmpsci}      % mathematics and physical sciences
\bibliographystyle{spphys}       % APS-like style for physics
\bibliography{sample-base}   % name your BibTeX data base

\end{document}